\documentclass[12pt,english]{article}
\pdfoutput=1
\usepackage[T1]{fontenc}
\usepackage{amssymb}
\usepackage{amsmath}
\usepackage{cite}
\usepackage{epsfig}
\usepackage{scalerel}
\usepackage[unicode=true,bookmarks=true,bookmarksnumbered=false,bookmarksopen=false,
 breaklinks=false,pdfborder={0 0 1},backref=false,colorlinks=true]
 {hyperref}
\usepackage{breakurl}
\usepackage[hang,flushmargin]{footmisc}
\usepackage{breakurl}
\usepackage{color}
\usepackage{shuffle}
\usepackage{cleveref}
\usepackage{nicematrix}

\usepackage[latin1]{inputenc}
\usepackage{amsmath}
\usepackage{amsfonts}
\usepackage{amssymb}

\usepackage{graphicx}
\usepackage[export]{adjustbox}

\usepackage[most]{tcolorbox}

\usepackage[bbgreekl]{mathbbol}
\DeclareMathSymbol\bbDelta  \mathord{bbold}{"01}

\makeatletter
\def\simgt{\mathrel{\lower2.5pt\vbox{\lineskip=0pt\baselineskip=0pt
           \hbox{$>$}\hbox{$\sim$}}}}
\def\simlt{\mathrel{\lower2.5pt\vbox{\lineskip=0pt\baselineskip=0pt
           \hbox{$<$}\hbox{$\sim$}}}}
\makeatother

\newcommand{\be}{\begin{equation}}
\newcommand{\ee}{\end{equation}}
\newcommand{\bea}{\begin{eqnarray}}
\newcommand{\eea}{\end{eqnarray}}
\newcommand{\eq}[2]{\be\begin{aligned}#1 \label{#2}\end{aligned}\ee}

\Crefname{equation}{Eq.}{Eqs.}

\ifdefined \Ref
\renewcommand{\Ref}[1]{Ref.~\cite{#1}}
\else
\newcommand{\Ref}[1]{Ref.~\cite{#1}}
\fi

\newcommand{\Fig}[1]{Fig.~\ref{#1}}
\newcommand{\Eq}[1]{Eq.~\eqref{#1}}
\newcommand{\Eqs}[1]{\Cref{#1}}
\newcommand{\Sec}[1]{Sec.~\ref{#1}}

\newcommand{\App}[1]{App.~\ref{#1}}

\newcommand{\ov}{\overline}

\newcommand{\vsp}{\phantom{\big|}}

\newcommand{\BAS}{{\scriptscriptstyle \textrm{BAS}}}
\newcommand{\GBAS}{{\scriptscriptstyle \textrm{GBAS}}}
\newcommand{\NLSM}{{\scriptscriptstyle \textrm{NLSM}}}
\newcommand{\YM}{ {\scriptscriptstyle\textrm{YM}} }
\newcommand{\GR}{{\scriptscriptstyle\textrm{GR}}}

\newcommand{\arrowNLSM}{\overset{ \vsp \NLSM \vsp}{\rightarrow}}
\newcommand{\arrowYM}{\overset{\vsp  {\scriptscriptstyle F^3}  \vsp }{\rightarrow}}

\newcommand{\qandq}{\qquad \textrm{and} \qquad}

\hypersetup{citecolor=blue,linkcolor=blue,urlcolor=blue}

  \newcommand\hl[1]{\tcbset{highlight math style={
    colframe=lightgray, colback=white!25!white,
    left=2mm, right=2mm, top=2mm, bottom=2mm
  }}\tcbhighmath{#1}}

  \newcommand\hlLOW[1]{\tcbset{highlight math style={
    colframe=lightgray, colback=white!25!white,
    left=2mm, right=2mm, top=0.25mm, bottom=2mm
  }}\tcbhighmath{#1}}
 
  \newcommand\hlHIGH[1]{\tcbset{highlight math style={
    colframe=lightgray, colback=white!25!white,
    left=2mm, right=2mm, top=2mm, bottom=0.25mm
  }}\tcbhighmath{#1}}
  
  \newcommand{\eqhl}[2]{
 \tcbset{highlight math style={
    colframe=lightgray, colback=white!25!white,
    left=4mm, right=4mm, top=2mm, bottom=4mm
  }} \be \tcbhighmath{ \begin{aligned}#1 \label{#2}\end{aligned} }\ee}

\numberwithin{equation}{section}

\begin{document}

\interfootnotelinepenalty=10000
\baselineskip=18pt

\hfill   CALT-TH-2021-029

\vspace{2.5cm}
\thispagestyle{empty}
\begin{center}
{\LARGE \bf
Covariant Color-Kinematics Duality \\
}
\bigskip\vspace{1cm}{
{\large Clifford Cheung and James Mangan}
} \\[7mm]
 {\it Walter Burke Institute for Theoretical Physics\\[-1mm]
    California Institute of Technology, Pasadena, CA 91125 } \let\thefootnote\relax\footnote{\noindent e-mail: \url{clifford.cheung@caltech.edu}} \\
 \end{center}
\bigskip
\centerline{\large\bf Abstract}
\begin{quote} \small

 We show that color-kinematics duality is a manifest property of the equations of motion governing currents and field strengths.  
For the nonlinear sigma model (NLSM), this insight enables an implementation of the double copy at the level of fields, as well as an explicit construction of the kinematic algebra and associated kinematic current.  As a byproduct, we also derive new formulations of the special Galileon (SG) and Born-Infeld (BI) theory.   

For Yang-Mills (YM) theory, this same approach reveals a novel structure---covariant color-kinematics duality---whose only difference from the conventional duality is that $1/\Box$ is replaced with covariant $1/D^2$.   Remarkably, this structure implies that YM theory is itself the covariant double copy of gauged biadjoint scalar (GBAS) theory and an $F^3$ theory of field strengths encoding a corresponding kinematic algebra and current.  Directly applying the double copy to equations of motion, we derive general relativity (GR) from the product of Einstein-YM and $F^3$ theory.  This exercise reveals a trivial variant of the classical double copy that recasts any solution of GR as a solution of YM theory in a curved background.

Covariant color-kinematics duality also implies a new decomposition of tree-level amplitudes in YM theory into those of GBAS theory.  Using this representation we derive a closed-form, analytic expression for all BCJ numerators in YM theory and the NLSM for any number of particles in any spacetime dimension.  By virtue of the double copy, this constitutes an explicit formula for all tree-level scattering amplitudes in YM, GR, NLSM, SG, and BI.

\end{quote}

\setcounter{footnote}{0}

\newpage

\setcounter{tocdepth}{2}

\tableofcontents    

\newpage

\section{Introduction}

Color-kinematics duality is an astonishing property of scattering amplitudes that links vastly disparate phenomena in nature: gravitation and the strong interactions.  
The earliest manifestations of this idea appeared in the pioneering work of Kawai, Lewellen, and Tye (KLT) \cite{KLT}, who derived explicit formulas relating closed and open string amplitudes.  Decades later, Bern, Carrasco, and Johansson (BCJ) \cite{BCJ1, BCJ2} beautifully generalized this notion to the domain of quantum field theory by expressing gravitational amplitudes as the ``square'' of gauge theory amplitudes.  In this construction, the mirrored structures of color and kinematics play a crucial role.
Since then, an intricate web of double copies has emerged, including biadjoint scalar (BAS) theory, Yang-Mills (YM) theory, general relativity (GR), the nonlinear sigma model (NLSM), the special Galileon (SG), and Born-Infeld (BI) theory.  See \cite{BCJReview} for a comprehensive review of the subject.

Color-kinematics duality is a mathematically indisputable fact governing the structure of scattering amplitudes.  But {\it why} is it true?  A proper answer to this question must not only explain why the double copy works, but also why it sometimes fails.
Alas, the underlying physical mechanism of this structure remains elusive, apart from some modest progress in the self-dual sector of YM theory \cite{DonalSDYM1,DonalSDYM2,DonalSDYM3,HenrikSDYM}
and the NLSM \cite{CliffXYZ}.

This paper is an attempt to elucidate color-kinematics duality beyond the context of amplitudes, instead appealing to the more prosaic tools of quantum field theory.   Here the ultimate aspiration might be to {\it derive} color-kinematics duality directly from the known textbook formulations of the double copy theories.  To this end we achieve partial progress: color-kinematics duality---or at least some variant of it---can be made {\it manifest} at the level of equations of motion provided one recasts the dynamics in terms of currents and field strengths rather than the traditional underlying degrees of freedom.  Armed with this understanding, we construct an explicit implementation of the double copy at the level of fields and equations of motion.  We also deduce the associated kinematic algebras, together with the corresponding currents whose conservation laws enforce the kinematic Jacobi identities.  As an application of our ideas we derive a closed-form expression for all tree-level BCJ numerators---and thus all tree-level scattering amplitudes---in YM, GR, NLSM, SG, and BI.  The outline of this paper is as follows.

We begin in \Sec{sec:BAS} by reviewing BAS theory and its ornamented cousin, gauged biadjoint scalar (GBAS) theory.  In our discussion we reiterate the well-known connection between equations of motion and tree-level scattering amplitudes, {\it i.e.}~Berends-Giele recursion \cite{BerendsGiele}.   This preamble will serve as a template for all subsequent analyses.  

In \Sec{sec:NLSM} we reformulate the NLSM in terms of the {\it chiral current}, whose dynamics are governed by the equation of motion in \Eq{EOM_NLSM}.  Remarkably, this description exhibits manifest color-kinematics duality, which is why its associated Feynman rules satisfy the kinematic Jacobi identities automatically.  With this understanding we construct the kinematic algebra of the NLSM in \Eq{kin_alg_NLSM} and recognize it as none other than the {\it diffeomorphism algebra}.  Afterwards, in \Eq{K_NLSM} we derive the kinematic current, whose conservation law enforces the kinematic Jacobi identities and which is equal, curiously, to the second derivative of the energy-momentum tensor.   Using \Eq{replace_NLSM}, we apply the double copy at the level of fields to obtain new  formulations of the SG in \Eqs{EOM_SG1,EOM_SG2,EOM_SG} and of BI theory in \Eqs{EOM_BI1,EOM_BI2}.

Pursuing an analogous strategy in \Sec{sec:YM}, we recast the dynamics of YM theory in terms of an equation of motion for the {\it field strength} in \Eq{EOM_YM}.    This setup exhibits a ``covariant color-kinematics duality'' that is formally identical to the standard duality except with the propagator $1/\Box$ replaced with covariant $1/D^2$.   Amazingly, we discover that YM theory is not irreducible but is in fact {\it itself} a ``covariant double copy'' of more primitive  building blocks: GBAS theory and a certain ``$F^3$ theory'' of field strengths.    
Implementing this covariant double copy at the level of fields with \Eq{replace_YM}, we then derive Einstein's equations from the equations of motion of Einstein-Yang-Mills (EYM) theory and $F^3$ theory.    An amusing corollary of this analysis is that any solution of GR has a dual interpretation as a solution of YM theory in a curved background.
Afterwards, we construct the kinematic algebra of $F^3$ theory in \Eq{kin_alg_YM} and realize that it is the {\it Lorentz algebra}, with generators given literally by the field strengths themselves, as shown in \Eqs{f_to_F,T_to_F}.  In \Eq{K_YM} we then derive the associated covariant kinematic current, which enforces the kinematic Jacobi identities appropriate to a theory with covariant propagators and also happens to be the first derivative of the energy-momentum tensor.

Covariant color-kinematics duality implies a new decomposition of amplitudes in YM theory and GR into those of GBAS and EYM theory times products of field strengths, as shown in \Eqs{field_strength_decomp,field_strength_decomp_grav}.     Happily, these representations can be utilized to derive \Eqs{result2,FandG,F_and_G_NLSM}, which are {\it fully analytic, closed-form} expressions for all BCJ numerators in YM theory and the NLSM for any number of external particles in arbitrary spacetime dimensions.   We emphasize that these formulas do not entail any implicit recursive definitions, unevaluated integrals, diagrammatic rules, or algorithmic prescriptions.  Perhaps surprisingly, these BCJ numerators are also manifestly gauge invariant and permutation invariant on all but one leg.   They also depend on arbitrary reference momenta which are easily chosen to generate BCJ numerators that are Lorentz invariant, local functions of the kinematics.  

With a formula for all BCJ numerators it is literally a matter of multiplication to derive the numerators for a multitude of other amplitudes via the standard double copy procedure.  Since all BAS amplitudes are known explicitly \cite{CHY2,MizeraPhiCubed}, our closed-form expressions constitute analytic formulas for all tree-level scattering amplitudes in YM, GR, NLSM, SG, and BI.

For completeness, we also present alternative formulations of the NLSM and YM theory in \App{app:twoform} and \App{app:cubic}.  The latter is a particularly compact description of YM theory defined solely in terms of a field strength, {\it i.e.} sans auxiliary fields, endowed with a single cubic self-interaction.  Last but not least, in \App{app:BCJ} we present a simple derivation of the fundamental BCJ relations using equations of motion.

\section{Biadjoint Scalar Theory}

\label{sec:BAS}

In this section we present a brief review of BAS theory which will function as a warmup for our later discussion of the NLSM and YM theory.  
 BAS theory describes a biadjoint scalar field $\phi^{a\ov a}$ with the Lagrangian,\footnote{We employ mostly minus metric conventions and index notation in which $ {\cal V} \overset{\leftrightarrow}{\partial}_\alpha  {\cal W} =  {\cal V} {\partial}_\alpha {\cal W} - {\cal W} {\partial}_\alpha {\cal V}$, while $\partial_{[\mu} {\cal V}_{\nu]} = \partial_{\mu} {\cal V}_{\nu}- \partial_{\nu} {\cal V}_{\mu}$ for a vector and $\partial_{[\rho} {\cal V}_{\mu \nu]} = \partial_{\rho} {\cal V}_{\mu\nu}  +  \partial_{\mu} {\cal V}_{\nu\rho} + \partial_{\nu} {\cal V}_{\rho\mu} $ for an antisymmetric tensor. We also use natural units in which all coupling constants are set to one.}
\eq{
{\cal L}^{\BAS} &= \tfrac{1}{2} \partial_\mu \phi^{a \ov a} \partial^\mu \phi^{a \ov a} - \tfrac{1}{3!}f^{abc} f^{\ov a \ov b \ov c} \phi^{a \ov a} \phi^{b \ov b} \phi^{c \ov c} + \phi^{a \ov a} J^{a \ov a}.
}{L_BAS}
Using the conventions of \cite{Parke}, we have introduced the structure constant $f^{abc}$ and generator $T^a$ of the color\footnote{We adopt an abuse of notation common amongst amplitudes practitioners in which ``color'' refers to any internal index, global or gauged. } algebra, which are related by
\eq{
{}[T^a, T^b] = i f^{abc} T^c \qquad \textrm{and} \qquad {\rm tr}\left[T^a T^b \right] =\delta^{ab},
}{normalizations}
and similarly for the dual color algebra encoded by $f^{\ov a\ov b \ov c}$ and $T^{\ov a}$.  
We assume throughout that the external source $J^{a \ov a}$ is localized at asymptotic infinity so as to produce on-shell external states. 

By construction, BAS theory is invariant under the global symmetry transformations
\eq{
\phi^{a\ov a} &\rightarrow  \phi^{a\ov a}  + f^{abc} \theta^b \phi^{c \ov a} \qandq
\phi^{a\ov a} &\rightarrow  \phi^{a\ov a}  + f^{\ov a \ov b \ov c}  \theta^{\ov b} \phi^{a \ov c},
}{}
for arbitrary constant parameters $\theta^a$ and $\theta^{\ov a}$.  The corresponding conserved currents are
\eq{
{\cal J}_\alpha^a &= f^{abc}  \phi^{b \ov a} \overset{\leftrightarrow}{\partial}_\alpha  \phi^{c \ov a}  \qandq
{\cal K}_\alpha^{\ov a} &=  f^{\ov a \ov b\ov c} \phi^{a \ov b}\overset{\leftrightarrow}{\partial}_\alpha \phi^{a \ov c} .
}{JK_BAS}
It is instructive to see how these currents are conserved on the support of the equations of motion.  In the absence of external sources, the divergence of the color current is
\eq{
\partial^\alpha {\cal J}_\alpha^a  &= f^{abc}  \phi^{b \ov a} \overset{\leftrightarrow}{\Box}  \phi^{c \ov a}  =  - f^{ade}  f^{ebc}  f^{\ov a \ov b\ov c} \phi^{d \ov a}\phi^{b \ov b} \phi^{c \ov c} = 0,
}{}
and similarly for the dual color current, for which $\partial^\alpha {\cal K}^{\ov a}_\alpha =0$.
Here the antisymmetric indices $\ov a \ov b \ov c$ contract into the indices of the scalars, thus imposing a cyclic symmetry on $bcd$.   The resulting expression is then proportional to the color Jacobi identity, $ f^{abe}  f^{ecd} +f^{ace}  f^{edb} +f^{ade}  f^{ebc} =0$.  
 
\subsection{Scattering Amplitudes}

As is well-known, tree-level scattering amplitudes are encoded in the solutions to equations of motion in the presence of an arbitrary source.  In other words, equations of motion are Berends-Giele recursion relations \cite{BerendsGiele} which implicitly define a set of Feynman rules that can be used to construct a perturbative solution or, alternatively, any tree-level scattering amplitude.   See \cite{MizeraEOM1} for earlier approaches investigating color-kinematics duality with equations of motion.

Let us review the mechanics of this procedure in the case of BAS theory.  We are interested in the perturbative solution of the BAS equation of motion, 
\eq{
\Box \phi^{a \ov a} + \tfrac{1}{2} f^{abc}  f^{\ov a \ov b\ov c} \phi^{b \ov b} \phi^{c \ov c} &= J^{a \ov a},
}{EOM_BAS}
expanded order by order in the source $J^{a \ov a}$.  The resulting solution $\langle \phi^{a\ov a}(p)\rangle_J$ is the one-point correlator in momentum space, which is also equal to the functional derivative of the connected partition function $W[J]$,
\eq{
\langle \phi^{a\ov a}(p)\rangle_J &= \frac{1}{i}\frac{\delta W[J]}{\delta J^{a \ov a}(p)} .
}{}
The $n$-point correlator is obtained from $n-1$ functional derivatives of the one-point correlator,
\eq{
\langle \phi^{a_1\ov a_1}(p_1)\phi^{a_2\ov a_2}(p_2) \cdots \phi^{a_n\ov a_n}(p_n)\rangle_{J=0} &= \left[ \left( \prod_{i=1}^{n-1} \frac{1}{i}\frac{\delta}{\delta J^{a_i \ov a_i}(p_i)} \right) \langle \phi^{a_n\ov a_n}(p_n)\rangle_J \right]_{J=0}  .
}{Greens_function_BAS}
Diagrammatically, each $n$-point correlator describes $n-1$ sources which propagate and fuse according the equations of motion, ultimately terminating at a single field which is the argument of the original one-point correlator. 
Here and throughout, we choose a convention in which the $n$-th leg of the $n$-point correlator is that field.   On account of the tree structure of this correlator, we refer to this $n$-th leg as the ``root'' leg and the other $n-1$ legs as the ``leaf'' legs.

The Feynman rules for BAS theory can be trivially derived by inspection from the equations of motion in \Eq{EOM_BAS}.  The propagator is
\eq{
\raisebox{0ex}{\includegraphics[trim={0 0 0 0},clip,valign=c,scale=0.8]{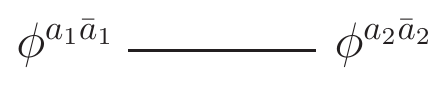}} &= \frac{i \delta^{a_1a_2} \delta^{\ov a_1 \ov a_2}}{p^2},
}{prop_BAS}
while the cubic interaction vertex is
\eq{
\raisebox{0ex}{\includegraphics[trim={0 0 0 0},clip,valign=c,scale=0.8]{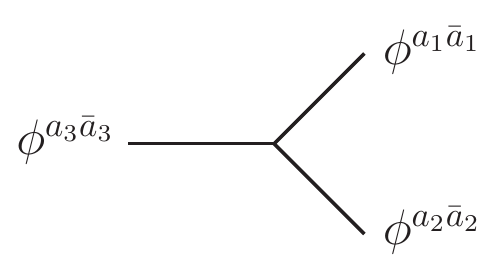}} &= - i f^{a_1 a_2 a_3} f^{\ov a_1 \ov a_2 \ov a_3} .
}{vert_BAS}
To compute the $n$-point correlator we simply sum over all Feynman diagrams connecting $n-1$ leaf legs to the root leg.  To obtain the $n$-point scattering amplitude we then amputate all external propagators.

As defined in \Eq{Greens_function_BAS}, the $n$-point correlator exhibits manifest permutation invariance on $n-1$ leaf legs.  However, Bose symmetry enforces full permutation invariance on all $n$ legs, so the root and leaf legs are in actuality interchangeable.  This only happens because the BAS equation of motion is derived from a Lagrangian whose {\it sole} degree of freedom is the scalar.  In general this is not guaranteed: that is, not every equation of motion can be derived from a Lagrangian whose only degrees of freedom are those already visible in the equations of motion.  In fact, this is possible if and only if the equations of motion satisfy a set of Helmholtz integrability conditions \cite{Helmholtz}.  Later on, we will encounter theories whose equations of motion simply fail these conditions.  In such circumstances the associated Feynman rules and $n$-point correlators are manifestly permutation invariant on the leaf legs but not the root leg.

\subsection{Gauged Formulation}

Last but not least we define GBAS theory, which is simply BAS theory with the color symmetry gauged.  The scalar sector of this theory is described by the Lagrangian,
\eq{
{\cal L}^{\GBAS} &= \tfrac{1}{2} D_\mu \phi^{a \ov a} D^\mu \phi^{a \ov a} - \tfrac{1}{3!}f^{abc} f^{\ov a \ov b \ov c} \phi^{a \ov a} \phi^{b \ov b} \phi^{c \ov c} + \phi^{a \ov a} J^{a \ov a},
}{L_GBAS}
where $D_\mu \phi^{a\ov a} = \partial_\mu \phi^{a \ov a} + f^{abc} A_\mu^b \phi^{c\ov a}$, so the dual color it not gauged.\footnote{GBAS theory is equivalent to the single trace sector of the ``YM + $\phi^3$ theory'' of \cite{Chiodaroli:2014xia,HenrikYMS}, {\it i.e.}~sans the quartic scalar interaction which necessarily enters at double trace order or higher.}
The equation of motion for GBAS theory is
\eq{
D^2 \phi^{a \ov a} + \tfrac{1}{2} f^{abc}  f^{\ov a \ov b\ov c} \phi^{b \ov b} \phi^{c \ov c} &= J^{a \ov a}.
}{EOM_GBAS}
The scalar propagator and self-interactions in GBAS theory are the same as in BAS theory.  On the other hand there are of course additional interactions involving the gauge field which we do not bother recapitulating here.  GBAS theory exhibits the conserved currents
\eq{
{\cal J}_\alpha^a &= f^{abc}  \phi^{b \ov a} \overset{\leftrightarrow}{D}_\alpha  \phi^{c \ov a}  \qandq
{\cal K}_\alpha^{\ov a} &=  f^{\ov a \ov b\ov c} \phi^{a \ov b}\overset{\leftrightarrow}{D}_\alpha \phi^{a \ov c} ,
}{JK_GBAS}
which are exactly the same as in \Eq{JK_BAS} except with covariant derivatives rather than partial derivatives.  Since color is gauged but dual color is not, the associated conservation equations are $D^\alpha {\cal J}_\alpha^a =0$ and $\partial^\alpha {\cal K}_\alpha^{\ov a} =0$, respectively.

\section{Nonlinear Sigma Model}
\label{sec:NLSM}

We are now equipped to study the NLSM.   The textbook formulation of this theory revolves around the traditional and well-studied Lagrangian for the NLSM scalar.  Here we instead reframe the dynamics of the NLSM in terms of the chiral current.

\subsection{Equations of Motion}

To begin, let us introduce an adjoint vector field $j^a_\mu$ with vanishing field strength, so
\eq{
\partial_{[\mu} j_{\nu]}^a + f^{abc} j_\mu^b j_\nu^c &=0 .
}{EOM_NLSM1}
This condition implies that the vector is a pure gauge configuration, so 
\eq{
j_\mu = j^a_\mu T^a = i g^{-1} \partial_\mu g,
}{pure_gauge}
where $g$ is an element of the color group which will ultimately encode the scalar field of the NLSM.  From this viewpoint, $j^a_\mu$ is nothing more than the chiral current of the NLSM.  
With this in mind we also impose the equation of motion for the NLSM,
\eq{
\partial^\mu j_\mu^a &= J^a,
}{EOM_NLSM2}
which says that the chiral current is conserved up to insertions of an external source $J^a$ that generates on-shell NLSM scalars at asymptotic infinity.  

\Eqs{EOM_NLSM1,EOM_NLSM2} comprise a first-order formulation of the NLSM.\footnote{To be precise, this  setup describes the NLSM of a {\it symmetric} coset space whose structure constants automatically satisfy the Jacobi identities and whose amplitudes exhibit the Adler zero condition \cite{Low:2014nga,Cheung:2020tqz}. }   A similar starting point was adopted in \cite{Freedman}, which proposed a novel representation of the NLSM in terms of the chiral current and an additional auxiliary antisymmetric tensor field.  By integrating out the latter those authors reproduced the canonical textbook Lagrangian for the NLSM.  Here we pursue a different strategy and {\it do not} attempt to reproduce any particular Lagrangian formulation, for several compelling reasons.  First of all, the theory space of putative auxiliary field completions is unbounded, so without any underlying guiding principles this exercise is infinitely open-ended.  Secondly, we need only reproduce the on-shell dynamics, so matching to any particular Lagrangian is actually highly over-constraining.  This is true because Lagrangians are inherently off-shell, nonunique objects, freely transformed via field redefinitions and integration by parts identities without altering on-shell observables.  For these reasons we opt to instead manipulate the equations of motion directly rather than reverse engineer a particular Lagrangian.

With this in mind we use \Eqs{EOM_NLSM1,EOM_NLSM2} to derive an equation of motion for the chiral current itself.  The combination of equations $\partial^\mu \left[\textrm{\Eq{EOM_NLSM1}}\right]_{\mu\nu} + \partial_\nu \left[\textrm{\Eq{EOM_NLSM2}}\right]  $ yields\footnote{Of course, the derivative of an equation admits spurious solutions which will no longer be valid solutions of the original equation.  However, we  avoid these pathologies when we solve perturbatively from the free theory.  }
\eq{
\hl{
\Box j_\mu^a  + f^{abc} j^{b\nu} \partial_\nu j^c_\mu = \partial_\mu J^a.
}
}{EOM_NLSM}
Here we have discarded nonlinear terms which are simultaneously proportional to both the source and the chiral current.   As is well-known, such couplings have no influence on on-shell scattering because the corresponding external sources are localized at asymptotic infinity, where fields linearize.  Indeed, this is precisely why on-shell scattering amplitudes are invariant under field redefinitions in the first place.  

The NLSM equation of motion in  \Eq{EOM_NLSM} describes a dynamical chiral current exhibiting a single cubic self-interaction and sourced by the {\it derivative} of the original NLSM scalar source.     It bears an uncanny resemblance to the equation of motion of BAS theory in \Eq{EOM_BAS}, and is in fact {\it identical} to that of the ``colored fluid'' \cite{NavierStokes}---theories which both, notably, exhibit manifest color-kinematics duality.
This observation will play an important role later on.

Last but not least, we note the existence of an alternative description of the NLSM in which the chiral current is dualized to an antisymmetric tensor field.  For more details, see \App{app:twoform}.

\subsection{Asymptotic States}

An immediate confusion now arises.  Even though we have formulated the NLSM in terms of the chiral current rather than the scalar, we ultimately care about the scattering of the latter and not the former.  How are the correlators of the NLSM scalars related to those of the chiral currents?  Naively, to answer this question we  need an explicit formula for the NLSM scalar in terms of the chiral current.  However, \Eq{EOM_NLSM1} only {\it implicitly} defines \Eq{pure_gauge}, so the precise field basis of the NLSM scalar is actually ambiguous.  Said another way, \Eq{EOM_NLSM1} does not actually specify the precise mapping between $g$ and the NLSM scalar field, 
\eq{
\pi = \pi^a T^a .
}{}
So for example, $g$ might be in the exponential basis, $g = e^{i\pi}$, or the Cayley basis, $g = \tfrac{1+i \pi/2}{1-i\pi/2}$.
 
For the purposes of calculating on-shell amplitudes it is actually {\it unnecessary} to specify this field basis.  To understand why, simply expand \Eq{pure_gauge} perturbatively in the NLSM field,
\eq{
j_\mu^a = - \partial_\mu \pi^a+ \cdots ,
}{}
where the ellipses denote terms that are nonlinear in the fields.  To invert this equation we introduce an arbitrary reference momentum $q$ and contract it with both sides to obtain
\eq{
\pi^a = -\frac{q^\mu j_\mu^a}{q\partial} + \cdots .
}{j_pi_relation}
For generic $q$, the inverse derivative operator $1/q\partial$ is well-defined. Crucially, the identities of the on-shell degrees of freedom are fully dictated by the linearized equations of motion.  Thus if the field is on-shell then the nonlinear terms in ellipses can be dropped.  In this case \Eq{j_pi_relation} implies that an on-shell NLSM scalar is equivalent to a peculiar reference-dependent polarization of an on-shell chiral current.

In perfect analogy with BAS theory, we then construct the scattering amplitudes of the NLSM by perturbatively solving the equation of motion for the chiral current in \Eq{EOM_NLSM} in the presence of sources.  From \Eq{j_pi_relation} we learn that the one-point correlators of the chiral current and the NLSM scalar are related by
\eq{
  \langle \pi^{a}(p)\rangle_J &= \tilde\varepsilon^\mu(p) \langle j^{a}_{\mu}(p)\rangle_J,
}{ej}
where the polarization of the root leg is
\eq{
\tilde \varepsilon_\mu(p) &= \frac{i q_\mu}{pq}.
}{e_root_NLSM}
Note again that we have assumed that the root leg is on-shell so that the nonlinear terms in \Eq{j_pi_relation} can be disregarded.  
Meanwhile, the $n$-point correlator is given by
\eq{
\langle \pi^{a_1}(p_1)\pi^{a_2}(p_2) \cdots \pi^{a_n}(p_n)\rangle_{J=0} &= \left[ \left( \prod_{i=1}^{n-1} \frac{1}{i}\frac{\delta}{\delta J^{a_i }(p_i)} \right) \tilde\varepsilon^\mu(p_n) \langle j^{a_n}_\mu(p_n) \rangle_J \right]_{J=0}  ,
}{Greens_function_NLSM}
where the functional derivatives are with respect to the NLSM scalar source.  In conclusion, one can extract the $n$-point correlator of NLSM scalars from the one-point correlator of the chiral current in the presence of sources.

\subsection{Kinematic Algebra}

The equations of motion for BAS theory and the NLSM in \Eqs{EOM_NLSM,EOM_BAS} are structurally identical.  By comparing them side by side we can derive the kinematic algebra {\it by inspection}.  In particular, BAS theory is mapped to the NLSM via three simple replacement rules, 
\eqhl{
{\cal V}^{ a} \quad &\arrowNLSM \quad {\cal V}_\mu \\
f^{ a  b  c} {\cal V}^{b} {\cal W}^{ c} \quad &\arrowNLSM \quad 
{\cal V}^\nu \partial_\nu {\cal W}_\mu - {\cal W}^\nu \partial_\nu {\cal V}_\mu\\
J^{a} \quad &\arrowNLSM \quad \partial_\mu J ,
}{replace_NLSM}
 which substitute color for kinematics.  First, we send any color index to a spacetime index.
 Second, we map color structure constants to kinematic structure constants.  Third, we replace any color sources with the derivative of a source.
Any indices unrelated to color---for instance those corresponding to the dual color---should be treated as spectator labels, left untouched.

The above replacement rules substitute the color algebra of any theory with the kinematic algebra of the NLSM, thus implementing the double copy at the level of fields.   For this reason we dub \Eq{replace_NLSM} the ``$\otimes$ NLSM replacement rules''.   This formulation of the double copy can be applied directly at the level of the fields in the equations of motion or, equivalently, at the level of Feynman rules derived from those equations of motion. 


The kinematic algebra of the NLSM is literally the diffeomorphism algebra.  This is confirmed by computing the commutator of generators,
\eq{
\hl{
{}[ \, {\cal V}_\mu \partial^\mu, {\cal W}_\nu \partial^\nu \, ] = ({\cal V}^\nu \partial_\nu {\cal W}_\mu - {\cal W}^\nu \partial_\nu {\cal V}_\mu) \partial^\mu,
}
}{kin_alg_NLSM}
which exactly reproduces the kinematic structure constants in \Eq{replace_NLSM}.  An identical construction of the kinematic algebra also appeared in the colored fluid of \cite{NavierStokes}.  It would be interesting to explore the precise connection between the diffeomorphism algebras which appear here and in self-dual YM theory \cite{DonalSDYM1,DonalSDYM2,DonalSDYM3,HenrikSDYM}.

\subsection{Double Copy}

\subsubsection{BAS $\otimes$ NLSM = NLSM}

It is illuminating to see how the procedure described above mechanically implements the double copy of BAS theory with the NLSM.    
To accomplish this we apply the $\otimes$ NLSM replacement rules to the dual color indices of BAS theory.  In particular, \Eq{replace_NLSM} sends the biadjoint scalar to the chiral current via
\eq{
\phi^{a \ov a} \quad \arrowNLSM \quad j^a_\mu,
}{}
where the spectator color index is unaffected by the replacement.  \Eqs{replace_NLSM} also sends  
\eq{
\Box \phi^{a \ov a} + \tfrac{1}{2} f^{abc}  f^{\ov a \ov b\ov c} \phi^{b \ov b} \phi^{c \ov c} = J^{a\ov a} \quad \arrowNLSM \quad \Box j_\mu^a  + f^{abc} j^{b\nu} \partial_\nu j^c_\mu  = \partial_\mu J^a,
}{}
thus deriving the equations of motion of the NLSM from those of BAS theory.

Any mapping between equations of motion is equivalent to a mapping between the Feynman rules derived from those equations of motion.   In particular, the $\otimes$ NLSM replacement rules in \Eq{replace_NLSM} send the propagator of BAS theory in \Eq{prop_BAS} to that of the NLSM,
\eq{
\raisebox{0ex}{\includegraphics[trim={0 0 0 0},clip,valign=c,scale=0.8]{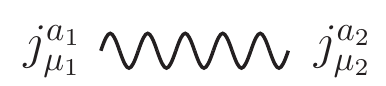}} &= \frac{i \delta^{a_1 a_2} \eta^{\mu_1\mu_2}}{p^2},
}{prop_NLSM}
and the Feynman vertex of BAS theory in \Eq{vert_BAS} to that of the NLSM,
\eq{
\raisebox{0ex}{\includegraphics[trim={0 0 0 0},clip,valign=c,scale=0.8]{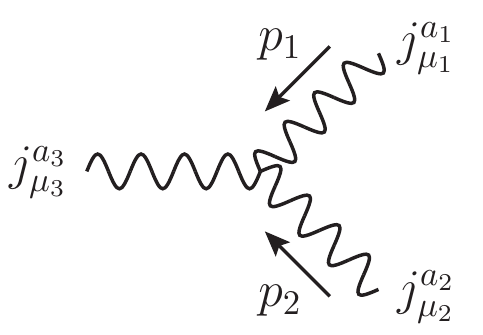}} &= - i f^{a_1 a_2 a_3} ( i p_2^{\mu_1} \eta^{\mu_2 \mu_3}-  i p_1^{\mu_2} \eta^{\mu_1\mu_3}).
}{vert_NLSM}
Curiously, the NLSM interaction vertex is only permutation invariant on legs 1 and 2, while leg 3 is actually special.  This is peculiar, but as noted previously it is also perfectly consistent because the NLSM equation of motion in \Eq{EOM_NLSM} cannot originate from a Lagrangian that depends solely on the chiral current.

Last of all we consider the external polarizations.  The $\otimes$ NLSM replacement rule in \Eq{replace_NLSM} sends the leaf leg polarizations of BAS theory---which are normally trivial---to
\eq{
\raisebox{0ex}{\includegraphics[trim={0 0 0 0},clip,valign=c,scale=0.8]{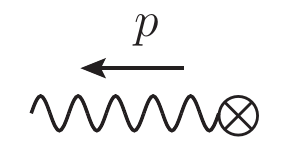}} = \varepsilon_\mu(p) &= i p_\mu,
}{pol_NLSM}
which are longitudinal sources for the chiral current of the NLSM.   As described earlier, the root leg polarization of the NLSM is
\eq{
\raisebox{0ex}{\includegraphics[trim={0 0 0 0},clip,valign=c,scale=0.8]{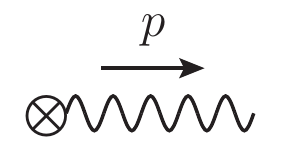}} = \tilde \varepsilon_\mu(p) &= \frac{i q_\mu}{pq},
}{root_NLSM}
which is needed to map the one-point correlator of the chiral current to that of the NLSM scalar.  

\Eqs{prop_NLSM,vert_NLSM,pol_NLSM,root_NLSM} define a new set of Feynman rules for the NLSM, and we have verified that they correctly reproduce all amplitudes up to eight-point scattering.  Moreover, the resulting Feynman diagrams {\it automatically} satisfy the kinematic Jacobi identities.  This can be proven simply by computing the {\it off-shell} four-point subdiagram of chiral currents embedded inside an arbitrary correlator.  Here legs 1,2, and 3 denote off-shell chiral currents that siphon into the remainder of the diagram.   Each of these external legs can be interpreted as a placeholder for some chiral current that will continually branch into others via the nonlinear self-interactions in the equations of motion, ultimately terminating at the leaf legs.
Meanwhile, we define leg 4 to be the root leg of this subdiagram, also taken to be off-shell.  The four-point subdiagram receives contributions from Feynman diagrams in the $s$, $t$, and $u$ channels,
\eq{
\raisebox{0ex}{\includegraphics[trim={0 0 0 0},clip,valign=c,scale=0.8]{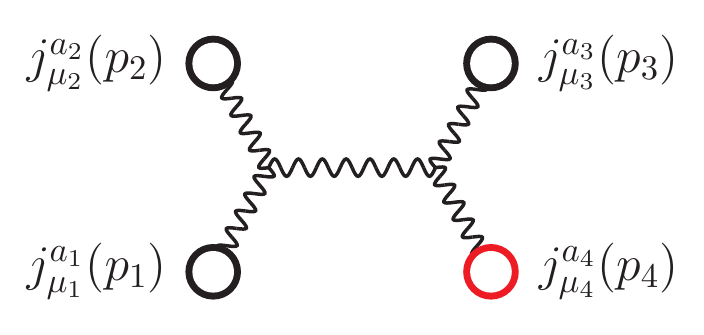}} &= \frac{c_s n_s }{s} \\ \\
\raisebox{0ex}{\includegraphics[trim={0 0 0 0},clip,valign=c,scale=0.8]{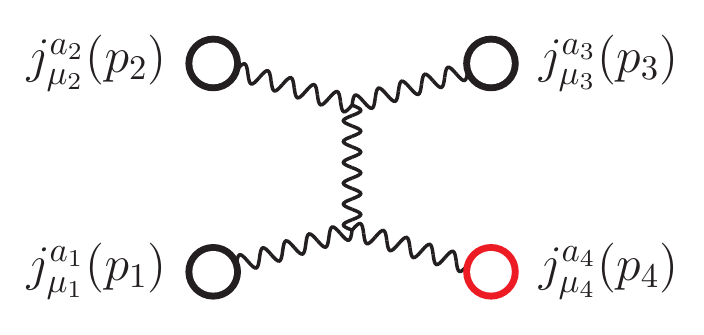}} &= \frac{c_t n_t }{t} \\ \\
\raisebox{0ex}{\includegraphics[trim={0 0 0 0},clip,valign=c,scale=0.8]{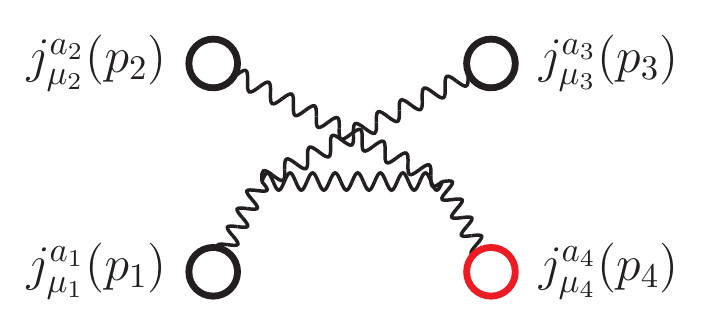}} &= \frac{c_u n_u }{u} ,
}{stu_diagrams}
where the circular blobs represent the off-shell chiral currents flowing to the rest of the diagram.
Note that the root leg, depicted in red, defines an orientation for the diagram which is important because the Feynman vertex in \Eq{vert_NLSM} treats the root leg as special.
The color structures of the $s$, $t$, and $u$ channel diagrams are
\eq{
c_s = f^{a_1 a_2 b} f^{b a_3 a_4}, \qquad c_t = f^{a_2 a_3 b} f^{b a_1 a_4}, \qquad c_u = f^{a_3 a_1 b} f^{b a_2 a_4},
}{}
and $c_s + c_t + c_u=0$ by the color Jacobi identity.
Using the Feynman rules in \Eqs{prop_NLSM,vert_NLSM,pol_NLSM,root_NLSM}, we obtain the kinematic numerators, 
\eq{
n_s &= p_1^{\mu_2} (p_1+p_2)^{\mu_3} \eta^{\mu_1\mu_4} +p_2^{\mu_1} p_3^{\mu_2} \eta^{\mu_3\mu_4} - \{ 1 \leftrightarrow 2\} \\
n_t &= p_2^{\mu_3} (p_2+p_3)^{\mu_1} \eta^{\mu_2\mu_4} + p_3^{\mu_2} p_1^{\mu_3} \eta^{\mu_1\mu_4} - \{ 2 \leftrightarrow 3\} \\
n_u &= p_3^{\mu_1} (p_3+p_1)^{\mu_2} \eta^{\mu_3\mu_4} + p_1^{\mu_3} p_2^{\mu_1} \eta^{\mu_2\mu_4} - \{ 3 \leftrightarrow 1\} ,
}{}
which are trivially related to each other by relabeling and algebraically satisfy
\eq{
n_s + n_t+n_u &= 0 .
}{kinematic_Jacobi_NLSM}
Hence, the kinematic Jacobi identities are automatically satisfied by the Feynman rules defined in \Eqs{prop_NLSM,vert_NLSM,pol_NLSM,root_NLSM}.   As a check, we have verified this claim up to eight-point scattering by explicit calculation.  These diagrammatic features reflect the underlying color-kinematics duality of the NLSM equations of motion in \Eq{EOM_NLSM}.

\subsubsection{NLSM $\otimes$ NLSM = SG}

The double copy of the NLSM is the SG, which was first discovered in the context of scattering amplitudes \cite{EFTFromSoft, CHY4} and then later understood in terms of symmetry \cite{Kurt_sGal}.  To implement this construction we apply the $\otimes$ NLSM replacement rules to the color indices of the NLSM.   \Eq{replace_NLSM} then sends the chiral current to
\eq{
j^{a }_\mu \quad \arrowNLSM \quad j_{\mu \ov \mu},
}{}
which we dub the ``chiral tensor''.  Now consider the first-order formulation of the NLSM, which is defined by the
vanishing field strength condition in \Eq{EOM_NLSM1}, together with the conservation equation in \Eq{EOM_NLSM2}.    \Eq{replace_NLSM} maps the field strength condition of the NLSM to
\eq{
\partial_{[\mu} j_{\nu]}^a + f^{abc} j_\mu^b j_\nu^c = 0 \quad & \arrowNLSM\quad  \partial_{[\mu} j_{\nu] \ov \mu} + 
j_\mu^{\;\;\ov\nu} \partial_{\ov \nu} j_{\nu \ov\mu}- j_\nu^{\;\;\ov\nu} \partial_{\ov \nu} j_{\mu \ov\mu}
=0 ,
}{EOM_SG1}
which we interpret as the statement that the generalized field strength of the chiral tensor is zero.  Meanwhile, the conservation equation of the NLSM is sent to
\eq{
\partial^\mu j_\mu^a = J^a  \quad & \arrowNLSM\quad  \partial^\mu j_{\mu\ov \nu} =  \partial_{\ov \nu} J,
}{EOM_SG2}
so the chiral tensor is also conserved, modulo external sources.

\Eqs{EOM_SG1,EOM_SG2} define a new first-order formulation of the SG theory.  In this description the scalar of the SG emerges as a pure gauge, longitudinal configuration of the chiral tensor.  This beautifully mirrors what happens in the NLSM, where the scalar emerges as a pure gauge, longitudinal configuration of the chiral current.  It would be interesting to explore whether this new formulation offers any new geometric insight into the SG theory or has any direct connection to massive gravity.

Taking the combination $\partial^\nu \left[\textrm{\Eq{EOM_SG1}}\right]_{\mu\nu \ov \mu} + \partial_\mu \left[\textrm{\Eq{EOM_SG2}}\right]_{\ov \mu}  $, we obtain a new equation of motion for the SG theory, 
\eq{
\Box j_{\mu \ov \mu}  + 
j^{\nu \ov \nu} \partial_\nu  \partial_{\ov \nu} j_{\mu \ov \mu}  -   \partial^\nu j_{\mu \ov \nu} \partial^{\ov \nu}  j_{\nu \ov \mu} = \partial_\mu \partial_{\ov \mu} J,
}{EOM_SG}
which can also be obtained more directly by applying the $\otimes$ NLSM replacement rules to the NLSM equation of motion in \Eq{EOM_NLSM}.

It is straightforward to derive the Feynman rules for the SG directly from \Eq{EOM_SG}. By construction they are the literal square of the Feynman rules of the NLSM.    Concretely, the propagator and interaction vertex of the SG are given by \Eqs{prop_NLSM,vert_NLSM}, except with all color structures replaced with yet another factor of the kinematic structures.  Meanwhile, the leaf and root leg polarizations of the SG are simply the squares of those in the NLSM shown in \Eqs{pol_NLSM,root_NLSM}.  We have verified that the resulting Feynman diagrams  reproduce the known amplitudes of the SG up to eight-point scattering.

\subsubsection{YM $\otimes$ NLSM = BI}

Next, let us implement the double copy of YM theory with the NLSM, which is BI theory.  To this end we apply the $\otimes$ NLSM replacement rules to the color indices of YM theory.  \Eq{replace_NLSM} maps the gauge field and field strength of YM theory to
\eq{
A^{a }_\mu \quad &\arrowNLSM \quad A_{\mu \ov \mu} \\
F^{a }_{\mu\nu} \quad &\arrowNLSM \quad F_{\mu \nu \ov \mu}.
}{}
As one might expect, the resulting tensor fields are not actually independent.  To see why, simply apply \Eq{replace_NLSM} to the relation between the gauge field and field strength of YM theory,
\eq{
F^a_{\mu\nu} = \partial_{[\mu} A^a_{\nu]} + f^{abc} A^b_\mu A^c_\nu \quad\arrowNLSM \quad F_{\mu\nu\ov\mu} = \partial_{[\mu} A_{\nu] \ov \mu} + 
A_\mu^{\;\;\ov\nu} \partial_{\ov \nu} A_{\nu \ov\mu}- A_\nu^{\;\;\ov\nu} \partial_{\ov \nu} A_{\mu \ov\mu}.
}{EOM_BI1}
Meanwhile, \Eq{replace_NLSM} also maps the YM equations of motion to
\eq{
\partial^\mu F_{\mu\nu}^a + f^{abc} A^{b\mu} F_{\mu\nu}^c= J_\nu^a \quad\arrowNLSM \quad  \partial^\mu F_{\mu\nu\ov\mu} + 
A^{\mu\ov\nu} \partial_{\ov \nu} F_{\mu \nu \ov\mu}- \partial_{\ov \nu} A^\mu_{\;\; \ov\mu} F_{\mu\nu}^{\;\;\;\;\ov\nu} 
= \partial_{\ov \mu} J_{\nu }.
}{EOM_BI2}
\Eqs{EOM_BI1,EOM_BI2} constitute a new first-order formulation of BI theory.  By construction, this representation of BI theory is structurally identical to YM theory, {\it e.g.}~the interaction vertices truncate at quartic order. A similar feature arose in the versions of BI theory described in \cite{CliffPionsAsGluons}.  

As before, it is trivial to derive the Feynman rules for BI theory from its equations of motion in \Eqs{EOM_BI1,EOM_BI2}.  The diagrammatics are a doppleganger of YM theory except with all color structure constants replaced with the kinematic numerators of the NLSM.  Going to the analog of Feynman gauge, we have verified up to six-point scattering that the resulting BI amplitudes agree with known expressions.

\subsection{Kinematic Current}

As reviewed in \Sec{sec:BAS}, the existence of a color algebra is inextricably linked to the conservation of a corresponding color current.  This current is conserved precisely as a consequence of the color Jacobi identities.

While the kinematic algebra is far more mysterious, we have at our disposal a new tool: an implementation of the double copy at the level of fields.  Applying the $\otimes$ NLSM replacement rule in \Eq{replace_NLSM} directly to the dual color current of BAS theory in \Eq{JK_BAS}, we obtain 
\eq{
\hlLOW{
{\cal K}_{\mu\alpha}^{\NLSM}   = j^{a\nu} \partial_\nu  \overset{\leftrightarrow}{\partial}_\alpha j^a_\mu,
}
}{K_NLSM}
which is the {\it kinematic} current of the NLSM.  
Taking the divergence of \Eq{K_NLSM}, we find that the kinematic current is conserved on the support of the NLSM equations of motion, so
\eq{
\partial^\alpha {\cal K}^{\NLSM}_{\mu \alpha} = j^{a\nu} \partial_\nu  \overset{\leftrightarrow}{\Box} j^a_\mu =  
-f^{abc} \left[    j^{a\nu} \partial_\nu (j^{b\rho} \partial_\rho j^c_\mu ) -  \partial_\nu j^a_\mu j^{b\rho} \partial_\rho j^{c\nu}   \right]  =0,
}{}
neglecting all external sources.  We emphasize that the conservation of the kinematic current should come as no surprise.  In fact, this is {\it mandated} in our formulation of the NLSM, since color-kinematics duality is manifest and the corresponding Feynman rules automatically satisfy the kinematic Jacobi identities.  

We also observe that the kinematic current is itself the derivative of yet another tensor,
\eq{
 {\cal K}_{\mu \alpha}^{\NLSM}  =  \partial^\nu  {\cal K}^{\NLSM}_{\mu \nu \alpha} \qquad \textrm{where} \qquad
  {\cal K}_{\mu \nu \alpha}^{\NLSM}  =  j^{a}_{\nu}  \overset{\leftrightarrow}{\partial}_\alpha j^a_\mu,
}{K2_to_K3}
where we have used the conservation equation for the chiral current in \Eq{EOM_NLSM2}.
Since ${\cal K}_{\mu\nu\alpha}^{\NLSM}$ is antisymmetric in its $\mu\nu$ indices, we learn that
\eq{
\partial^\mu {\cal K}_{\mu \alpha}^{\NLSM} =  \partial^\mu \partial^\nu  {\cal K}_{\mu \nu \alpha}^{\NLSM}=0,
}{K_conserved_NLSM}
so the kinematic current is conserved on {\it both} of its spacetime indices.  

The fact that the kinematic current is a {\it color singlet} tensor conserved on both indices is an important hint as to its true identity.
Perhaps unsurprisingly, the kinematic current is intimately related to the energy-momentum tensor of the NLSM,
\eq{
T_{\mu\nu}^{\NLSM} =j_\mu^a j_\nu^a -\tfrac{1}{2} \eta_{\mu\nu} j_\rho^a j^{a\rho}.
}{T_NLSM}
On the support of \Eq{EOM_NLSM1}, the tensor in \Eq{K2_to_K3} can be written as 
\eq{
{\cal K}_{\mu\nu\alpha}^{\NLSM}   = \partial_{[\mu} T_{\nu]\alpha}^{\NLSM} - \left(f^{abc}  j^a_\mu j^b_\nu j^c_\alpha +\tfrac{1}{2} \eta_{\alpha [\mu} \partial_{\nu]} (j_\rho^a j^{a\rho}) \right).
}{}
Notably, this quantity is {\it not} conserved because of the second term.  However, \Eq{K2_to_K3} then implies that the kinematic current is equal to
\eq{
{\cal K}_{\mu\alpha}^{\NLSM}   &= \partial^\nu \partial_{[\mu} T_{\nu]\alpha}^{\NLSM} -\partial^\nu \left( f^{abc}   j^a_\mu j^b_\nu j^c_\alpha + \tfrac{1}{2} \eta_{\alpha [\mu} \partial_{\nu]} (j_\rho^a j^{a\rho})\right) \\
& 
= -\Box T_{\mu\alpha}^{\NLSM} + \textrm{improvement terms},
}{K_NLSM_simp}
where we have discarded contributions that vanish by energy-momentum tensor conservation and set aside improvement terms that are trivially conserved by antisymmetry.   Thus, the kinematic current is nothing more than the second derivative of the energy-momentum tensor.

Is the kinematic current associated with a symmetry?   Strangely, the answer appears to be no.  In fact, this is a general feature of any current which is also the derivative of another current.  To understand why, consider a theory whose symmetries induce a conserved Noether current, $J_\alpha$.  
The associated conservation equation, $\partial^\alpha J_\alpha=0$, implies that the charge, $Q = \int d^3 x \, J_0(x)$, is constant in time, so $\partial_0 Q=0$.   Now consider the {\it derivative} of the Noether current, $\partial_\mu J_\alpha$, which is also conserved, albeit trivially.  Since $\partial^\alpha \partial_\mu J_\alpha=0$, we can also define a new charge, $Q_\mu = \int d^3 x\, \partial_\mu J_0(x)$, which is also a constant of motion, so $\partial_0 Q_\mu=0$.   This quantity is secretly zero, however.  One can see this via explicit calculation, since $Q_0 = \partial_0 Q = 0$ and  $Q_i=0$ is the volume integral of a total derivative in Cartesian coordinates.\footnote{Since $Q_\mu$ is the volume  integral of a tensor it is not actually coordinate invariant, {\it i.e.}~its value depends on the choice of coordinates.  This is certainly peculiar.  However, in any particular choice of coordinates it will still be a constant of motion and thus operationally useful for solving the initial value problem.  } 
 Physically, this triviality arises because $Q_\mu$ measures the {\it change} in $Q$ across an infinitesimal interval, which is to say that $a^\mu Q_\mu = \lim\limits_{a\rightarrow 0}\int d^3 x \,  \left[ J_0(x+a)- J_0(x) \right]$.  Since the integration surface is assumed to be translationally invariant and $Q$ is constant, the first and second terms exactly cancel.  In other words, if $Q$ is a constant of motion then so too is any infinitesimal change in $Q$---in which case the corresponding constant of motion also happens to be zero.   We thus conclude that the kinematic current, like any derivative of a conserved current, corresponds to a charge that acts trivially on all physical states.   Still, it would be interesting to explore whether the relationship between the kinematic current and the energy-momentum tensor has any link to the fact that scalar double copy theories are actually fixed by conformal invariance \cite{Cheung:2020qxc,Farnsworth:2021ycg}.

\section{Yang-Mills Theory}

\label{sec:YM}

Our analysis of the NLSM offers a clear roadmap for generalization to YM theory.  For the NLSM, we repackaged all of the dynamics in terms of the chiral current rather than the underlying scalar.  The analogous strategy for YM theory would be to reformulate the dynamics in terms of the field strength rather than the gauge field.  

Surprisingly, this is easily achievable.  In fact, it is trivial to derive equations of motion for YM theory in terms of the field strength alone, {\it i.e.}~sans auxiliary degrees of freedom.  Furthermore, in this representation the field strength interacts solely through a single cubic vertex.  While the resulting cubic formulation of YM theory is exceedingly compact, it does not exhibit any obvious manifestation of color-kinematics duality, and for this reason we relegate any discussion of it to \App{app:cubic}.

For the remainder of this section we pursue an alternative path which incorporates both the field strength {\it and} the gauge field, albeit with the latter entering far more mildly.  In particular, we derive an equation of motion for the field strength in which the gauge field enters solely through the kinetic term.  In this formulation the $1/\Box$ propagators that play such a central role in color-kinematics duality are formally replaced with covariant $1/D^2$ propagators.   This representation of YM theory exhibits a covariant version of color-kinematics duality that preserves gauge invariance at every step.    As we will see, this new duality implies that YM theory is itself a covariant double copy of more basic building blocks.

\subsection{Equations of Motion}

Parroting our earlier analysis of the NLSM, we adopt a first-order formulation of YM theory as our starting point.  In this description the gauge field $A_\mu^a$ and the field strength $F_{\mu\nu}^a$ are treated as a priori independent objects.  First, we assume the Bianchi identity,
\eq{
D_{[\rho} F_{\mu\nu]}^a&=0,
}{EOM_YM1}
where the covariant derivative acts as $D_\rho F^a_{\mu\nu} = \partial_\rho F^a_{\mu\nu} + f^{abc} A_\rho^b F^c_{\mu\nu}$.  Second, we impose the usual equations of motion of YM theory,
\eq{
D^\mu F_{\mu\nu}^a&= J_\nu^a,
}{EOM_YM2}
where $J_\mu^a$ is an external source for the gauge field.  As before, we stipulate that this source only generates on-shell quanta localized at asymptotic infinity.

Taking the combination of equations  $D^\rho \left[\textrm{\Eq{EOM_YM1}}\right]_{\rho\mu\nu} + D_{[\mu} \left[\textrm{\Eq{EOM_YM2}}\right]_{\nu]}  $, we derive an equation of motion for the field strength,
\eq{
\hl{
D^2 F^{a}_{\mu\nu} +f^{abc} F^{b}_{\rho [\mu}  F^{c\rho}_{\nu]} = D_{[\mu} J_{\nu]}^{a} ,
}
}{EOM_YM}
where we have used that $D_{[\rho} D_{\sigma]} F^a_{\mu\nu} = f^{abc} F_{\rho\sigma}^b F^c_{\mu\nu}$.  Notice that the gauge field only appears through covariant derivatives acting on the kinetic term and the source term.  However, we have assumed that $J^a_\mu$ is localized at asymptotic infinity, so all nonlinear field interactions involving the current can be dropped.  Hence, the $D_{[\mu} J_{\nu]}^{a}$ source term is, for the purposes of on-shell scattering, the same as $\partial_{[\mu} J_{\nu]}^{a}$, though we maintain the covariant form throughout.  

\begin{figure}
\begin{center}
\raisebox{0ex}{\includegraphics[trim={0 0 0 0},clip,valign=c,scale=1]{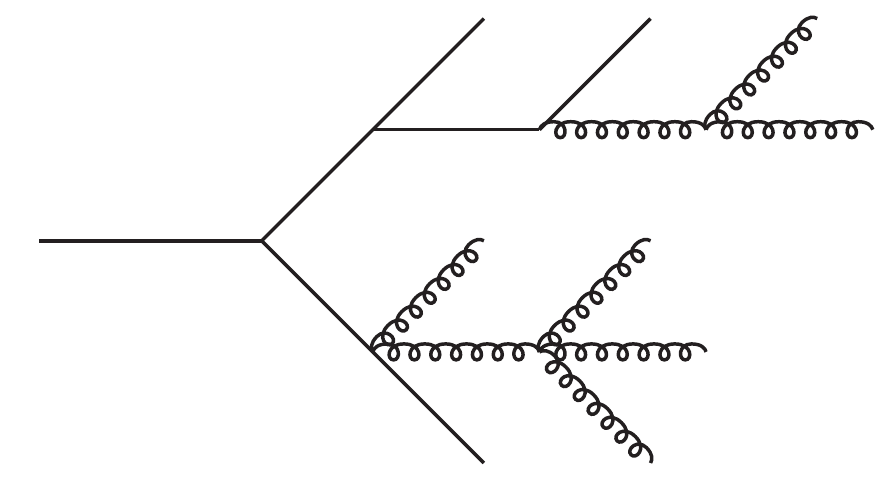}} 
\end{center}
\caption{A typical Feynman diagram that might appear in GBAS theory, where the solid and curly lines depict biadjoint scalars and gauge fields, respectively.    This has a dual interpretation in YM theory, where the solid and curly lines depict field strengths and gauge fields, respectively.  The root leg is depicted on the left, with all leaf legs to the right.
}
\label{fig:GBAS_YM}
\end{figure}

\Eq{EOM_YM} offers a new angle on the dynamics of YM theory.
The field strength is minimally coupled to the gauge field precisely like a charged scalar.  Furthermore, it exhibits a single cubic self interaction and is externally sourced by the derivative of the gauge field source.  Said another way, the equation of motion for YM theory in \Eq{EOM_YM} is structurally identical to that of GBAS theory in \Eq{EOM_GBAS}.  
 This is no accident.  It is the harbinger of covariant color-kinematics duality and has an elegant physical interpretation: field strengths evolve exactly like charged, self-interacting scalars.   Notably, this implies an isomorphism between the Feynman diagrams of YM and GBAS theory which are derived from their respective equations of motion.  See \Fig{fig:GBAS_YM} for an illustration of this connection.

\subsection{Asymptotic States}

The equation of motion in \Eq{EOM_YM} governs the dynamics of the field strength but our end goal is to understand the scattering amplitudes of the underlying gauge field.  Fortunately, we already encountered and the very same problem in the NLSM and solved it.  For YM theory, the analogous strategy is clear: rewrite the gauge field in terms of the field strength.  To this end, recall that the Bianchi identity in \Eq{EOM_YM1} is automatically satisfied when
\eq{
F^a_{\mu\nu} = \partial_{[\mu} A^a_{\nu]} + f^{abc} A^b_\mu A^c_\nu,
}{F_from_A}
which is the familiar relationship between the gauge field and field strength.  
Of course, it is only possible to invert \Eq{F_from_A} if we also fix a gauge.  For example, in axial gauge we have that
\eq{
q^\mu A^a_\mu &= 0,
}{}
in which case \Eq{F_from_A} can be solved to obtain
\eq{
A^a_\mu = - \frac{q^\nu F_{\mu \nu}^a}{q\partial} ,
}{A_from_F}
for an arbitrary reference vector $q$.  Thus, in axial gauge one can {\it entirely eliminate} the gauge field  in favor of the field strength.  We do precisely this in \App{app:cubic} in order to derive a purely cubic reformulation of YM theory in terms of a self-interacting field strength.

For the present discussion, however, we adopt a more agnostic approach that is purposely {\it noncommittal} to the precise choice of gauge fixing.  Our motivations here are the same as for the NLSM, where the nonlinear map connecting the NLSM scalar and the chiral current was not actually specified by the equations of motion.  Rather than designate an arbitrary choice there, we fixed the NLSM field basis unambiguously at linear order and simply discarded any residual nonlinear differences.  This was permitted because the nonlinearities in the map are field basis dependent and thus evaporate from physical on-shell scattering.  Those differences would have persisted had we instead computed off-shell correlators.

In the case of YM theory, the selection of axial gauge---or any gauge for that matter---is as arbitrary as fixing a particular field basis in the NLSM.   
For these reasons, we make the weaker assumption that the gauge field and field strength are related at linear order by \Eq{A_from_F}, but differ at nonlinear order by arbitrary contributions that depend on the gauge fixing.   Hence, the gauge field is on-shell equivalent to a particular exotic polarization of the field strength and consequently the one-point correlators are related by
\eq{
\langle A^{a}_\mu(p)\rangle_J &= \tilde\varepsilon^\nu(p) \langle F^{a}_{\mu \nu}(p)\rangle_J ,
}{F_to_A_corr}
provided the root leg is on-shell.  Note the similarity of this expression to \Eq{ej}.

Given the one-point correlator of the field strength we can extract the $n$-point correlator of the gauge field in the usual way via functional differentiation, 
\eq{
\langle A_{\mu_1}^{a_1}(p_1)A_{\mu_2}^{a_2}(p_2) \cdots A_{\mu_n}^{a_n}(p_n)\rangle_{J=0} &= \left[ \left( \prod_{i=1}^{n-1} \frac{1}{i}\frac{\delta}{\delta J^{a_i \mu_i}(p_i)} \right) \tilde\varepsilon^{\nu_n} (p_n) \langle F^{a_n}_{\mu_n \nu_n}(p_n) \rangle_J \right]_{J=0}  .
}{Greens_function_YM}
To summarize, any $n$-point correlator of gauge fields is encoded within the one-point correlator of the field strength in the presence of sources.  We emphasize again that when evaluated on-shell, \Eqs{F_to_A_corr,Greens_function_YM} are independent of any arbitrary choices of gauge fixing and field basis that appear at nonlinear order in the fields.

\begin{figure}
\begin{center}
\begin{tabular}{l l}
$A_3^{\YM} $ & = \raisebox{0.3ex}{\includegraphics[trim={0 0 0 0},clip,valign=c,scale=0.7]{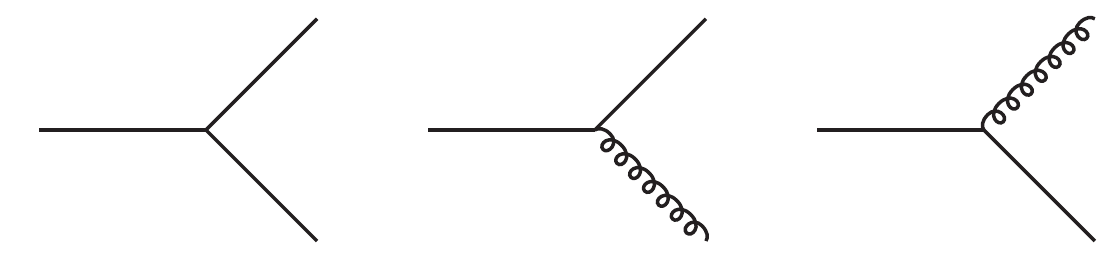}} \\
~&~\\
~&~\\
\raisebox{6.3ex}{$A_4^{\YM} $} & \raisebox{6.3ex}{=} \raisebox{0ex}{\includegraphics[trim={0 0 0 0},clip,valign=c,scale=0.7]{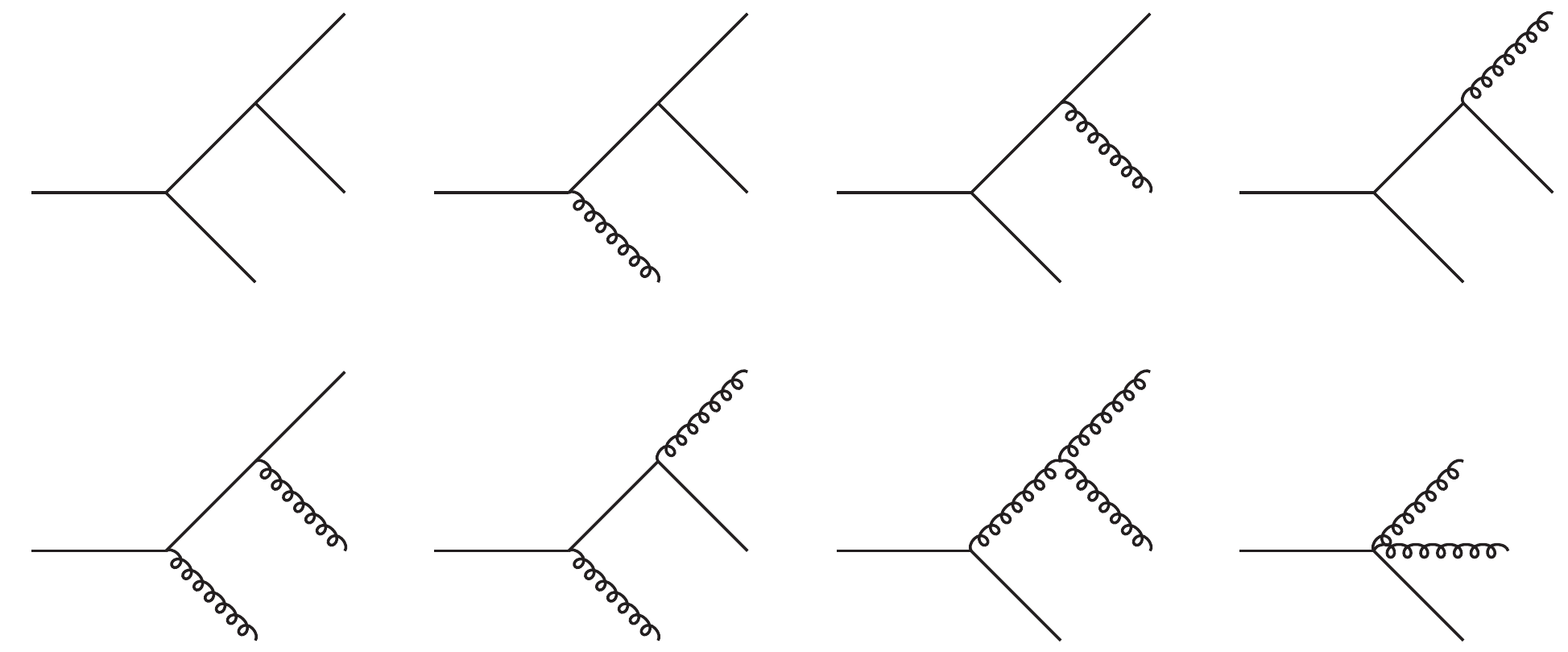}} \\
~&~\\ \\
 & ~~~~ + $t$-channel + $u$-channel
\end{tabular}
\end{center}
\caption{The three- and four-point scattering amplitudes of YM theory derived from the one-point correlator of the field strength in \Eq{Greens_function_YM} using the Feynman rules derived from the equation of motion in \Eq{EOM_YM}.  The solid and curly lines depict field strengths and gauge fields, respectively.  Embedded within in each diagram is a subdiagram of solid lines describing a field strength that branches purely through $F^3$ interactions.  Like minimally coupled scalars, these field strengths repeatedly emit gauge fields that in turn cascade down into other gauge fields via the nonlinear interactions of YM theory.   Since the gauge fields never branch back into field strengths, they are effectively background fields and induce no back-reaction.   Note that the root leg is always a field strength, while the leaf legs can be either gauge fields {\it or} field strengths, since these are generated at asymptotic infinity by the source and its derivative, respectively.
 }
\label{fig:4pt}
\end{figure}

To compute the right-hand side of \Eq{F_to_A_corr} one perturbatively solves \Eq{EOM_YM} to obtain the one-point correlator of the field strength as a functional of the external source.
An immediate confusion now arises because the perturbative solution for the field strength does not just depend on the external source.  It also depends on the gauge field, which enters  \Eq{EOM_YM} through covariant derivatives.   There is, however, no cause for concern.  We simply realize that the one-point correlator of the gauge field also evolves by its own equation of motion, and like the field strength should be solved for perturbatively.   Due to gauge symmetry, there is immense freedom in explicitly solving for this gauge field.  But the simplest procedure is to plug \Eq{F_from_A} directly into \Eq{EOM_YM2}, yielding the standard equations of motion for the gauge field of YM theory in the presence of a source.  The perturbative solution to this equation encodes the one-point correlator for the gauge field computed using standard Berends-Giele recursion \cite{BerendsGiele}.    In this prescription, gauge fields which are emitted by the field strength will never branch back into field strengths.  Hence, this one-point correlator should then be treated like a {\it background} gauge field to be inserted into  \Eq{EOM_YM} when solving for the field strength.   This is highly reminiscent of restricting to the sector of GBAS theory which is {\it single trace} in the dual color, and relates intimately to the covariant double copy structure we will discuss later on.
See \Fig{fig:4pt} for a representative Feynman diagram that would arise from this procedure.

The need to compute the one-point correlator of the gauge field would seem to defeat the purpose of recasting perturbation theory in terms of the one-point correlator of the field strength.  Indeed, the Feynman diagrams encoded in \Eq{EOM_YM} are, in practice, just as complicated as in standard perturbation theory, due to the appearance of the gauge field in the covariant derivatives.   In spite of this, however, we will see that organizing the equations of motion in this way will reveal a new structure within YM theory that will ultimately enable us to derive analytic expressions for all BCJ numerators.

\subsection{Kinematic Algebra}

The equations of motion for GBAS and YM in \Eqs{EOM_GBAS,EOM_YM} are isomorphic, with the former mapped onto the latter via three simple substitutions,
\eqhl{
V^{ a} \quad &\arrowYM \quad V_{\mu\nu} \\
f^{ a b c} {\cal V}^{ b} {\cal W}^{ c}\quad  &\arrowYM \quad  
{\cal V}_{\rho \mu}  {\cal W}^{\;\;\rho}_{\nu} - {\cal W}_{\rho \mu}  {\cal V}^{\;\;\rho}_{\nu}  \\
J^{a}\quad &\arrowYM \quad \partial_{[\mu} J_{\nu]},
}{replace_YM}
 which we dub the ``$\otimes \; F^3$ replacement rules'', for reasons which will soon become apparent.  First, we send any color index to a pair of antisymmetric spacetime indices.  Second, we substitute all color structure constants with kinematic structure constants.  Third, we send any color source to the antisymmetric derivative of a vector source.
 As before, any indices distinct from color are spectators to the above replacements.  Moreover, if any of these spectator indices are gauged, then the derivative on the right-hand side of the replacement rule for the source can be made covariant with impunity.  This is permitted because all external sources are localized at asymptotic infinity, where partial and covariant derivatives are interchangeable.

The kinematic structure constant in \Eq{replace_YM} is quite literally the Feynman vertex for a theory of cubically interacting, antisymmetric tensor fields.  This $F^3$ theory is the natural {\it tensorial generalization} of BAS theory, differing only in that dual color indices are replaced with pairs of antisymmetric spacetime indices.  Viewed purely as a description of generic antisymmetric fields, the $F^3$ theory is perfectly consistent, albeit trivial.  However, we will see how the plot thickens when we ultimately identify the antisymmetric fields as field strengths.  

Another alias for the $F^3$ vertex is the ``anomalous triple gauge boson coupling'', which traditionally appears as the leading {\it higher-derivative} correction to YM theory.  However, we stress emphatically that our discussion here pertains to the standard {\it renormalizable} interactions of YM theory.\footnote{The $F^3$ theory which has emerged here is similar but not identical to the dimension-six gauge theory which double copies to conformal gravity \cite{HenrikF3}.  It would be interesting to further explore this relationship.}  As we will see later on, the naive mismatch between numbers of derivatives is resolved because GBAS theory has fewer derivatives per interaction than YM theory while $F^3$ theory has more.  We dub \Eq{replace_YM} the $\otimes \; F^3$ replacement rules because it sends color structures to the kinematic structures of $F^3$ theory.

The substitutions described above are reminiscent of the usual double copy prescription of color-kinematics duality.  But they differ in a critical respect. The traditional formulation of the double copy instructs us to swap color for kinematics in a representation of the amplitude in which all propagators are of the form $1/\Box$.  By contrast, the equations of motion for GBAS and YM theory in \Eqs{EOM_GBAS,EOM_YM} carry kinetic terms with $D^2$ rather than $\Box$.    Thus, the $\otimes \; F^3$ replacement rules implement a covariant version of the double copy that applies to amplitudes whose propagators are effectively $1/D^2$.  Since gauge invariance is preserved by this procedure, we refer to this structure as covariant color-kinematics duality.  

Amusingly, the kinematic algebra of $F^3$ theory is actually the Lorentz algebra.  To understand why, we simply observe that \Eq{replace_YM} is exactly reproduced by the commutator
\eq{
\hl{
{} [ \, {\cal V}_{\mu\nu}  S^{\mu\nu}  ,  {\cal W}_{\rho \sigma}  S^{\rho\sigma} ] = ({\cal V}_{\rho \mu}  {\cal W}^{\;\;\rho}_{\nu} - {\cal W}_{\rho \mu}  {\cal V}^{\;\;\rho}_{\nu}  ) S^{\mu\nu},
}
}{kin_alg_YM}
where ${\cal V}_{\mu\nu}$ and ${\cal W}_{\mu\nu}$ are antisymmetric tensor fields parameterizing infinitesimal {\it spin Lorentz transformations} generated by $S^{\mu\nu}$, appropriately normalized.
These transformations are boosts and rotations which act on the spacetime indices of fields while leaving the coordinate arguments within those fields untouched.  This is why the right-hand side of \Eq{kin_alg_YM} does not include any derivatives of fields.
It would be interesting to understand if there is any link between the kinematic algebra described above and the spin Lorentz symmetries of \cite{NimaJared,TwofoldCliffGrant,Cheung:2017kzx}.

\subsection{Double Copy}

\subsubsection{GBAS $\otimes$ F${}^3$ = YM}

Covariant color-kinematics duality implies that YM theory is the covariant double copy of GBAS theory and $F^3$ theory.  \Eq{replace_YM} sends the biadjoint scalar to the field strength,
\eq{
\phi^{a \ov a} \quad\arrowYM\quad F^a_{\mu\nu},
}{}
and maps the equation of motion of GBAS theory to
\eq{
D^2 \phi^{a \ov a} + \tfrac{1}{2} f^{abc}  f^{\ov a \ov b\ov c} \phi^{b \ov b} \phi^{c \ov c}= J^{a \ov a} \quad \arrowYM \quad D^2 F^{a}_{\mu\nu} +f^{abc} F^{b}_{\rho [\mu}  F^{c\rho}_{\nu]} = D_{[\mu} J^a_{\nu]},
 }{GBAS_to_YM_EOM}
 which is precisely the YM equation of motion in \Eq{EOM_YM}.
 The physical interpretation of this result is that YM theory is secretly equivalent to GBAS theory but with the charged scalars dressed with additional spacetime indices.

Any feature of equations of motion is also a feature of the associated Feynman rules.  Thus \Eq{replace_YM} defines a map from the Feynman rules of GBAS theory to those of YM theory.
For example, the scalar propagator in \Eq{prop_BAS} is sent to the propagator for the field strength,
\eq{
\raisebox{0ex}{\includegraphics[trim={0 0 0 0},clip,valign=c,scale=0.8]{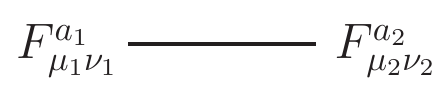}}  &= \frac{i \delta^{a_1a_2} \Pi^{\mu_1\nu_1\mu_2\nu_2}}{p^2} ,
}{prop_F3}
where we have defined the identity operator for antisymmetric tensor fields
\eq{
\Pi^{\mu_1\nu_1\mu_2\nu_2} &=\tfrac{1}{2}( \eta^{\mu_1\mu_2} \eta^{\nu_1\nu_2}- \eta^{\mu_1\nu_2} \eta^{\nu_1\mu_2}) .
}{}
Meanwhile, the cubic scalar vertex in \Eq{vert_BAS} is sent to
\eq{
\raisebox{0ex}{\includegraphics[trim={0 0 0 0},clip,valign=c,scale=0.8]{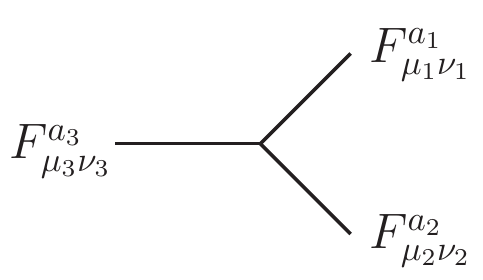}} &= 4 \, i f^{a_1a_2a_3} \Pi^{\mu_1\nu_1 \alpha}_{\;\;\;\;\;\;\;\;\;\; \beta} \Pi^{\mu_2\nu_2 \beta}_{\;\;\;\;\;\;\;\;\;\; \gamma}\Pi^{\mu_3\nu_3 \gamma}_{\;\;\;\;\;\;\;\;\;\; \alpha},
}{vertex_F3}
and the trivial leaf leg polarizations of scalars are mapped to field strength polarizations,
\eq{
\raisebox{0ex}{\includegraphics[trim={0 0 0 0},clip,valign=c,scale=0.8]{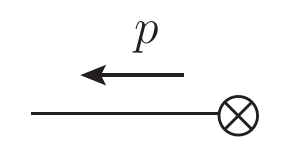}} =\varepsilon_{\mu\nu}(p) = i p_{[\mu} \varepsilon_{\nu]} .
}{leaf_pol_F3}
As discussed earlier the root leg polarization for the field strength is
\eq{
\raisebox{0ex}{\includegraphics[trim={0 0 0 0},clip,valign=c,scale=0.8]{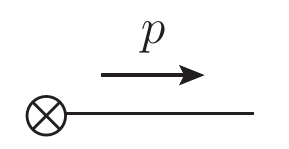}} = \tilde \varepsilon_{\mu\nu}(p) &= \frac{i q_{[\mu} \varepsilon_{\nu]}}{pq},
}{root_pol_F3}
as required to extract the one-point correlator of the gauge field from that of the field strength.

Crucially, the $\otimes \; F^3$ replacement rules trivially map all of the Feynman rules linking the scalar and gauge field of GBAS theory to those linking the field strength and gauge field of YM theory.  This is obvious from \Eq{GBAS_to_YM_EOM} because the gauge field appears solely in the kinetic terms, which are $D^2 \phi^{a\ov a}$ and $D^2 F^a_{\mu\nu}$.  Said another way, in this formulation the field strength couples to the gauge field exactly as if it were a minimally coupled scalar.  

This all implies that the Feynman rules of GBAS and YM theory are nearly isomorphic, modulo two important caveats.  First of all, as noted earlier, the field strengths in our formulation of YM theory emit gauge fields which can readily cascade into other gauge fields but never branch back into field strengths.  Consequently, YM amplitudes are mapped to amplitudes in GBAS theory which are single trace in dual color, so gauge bosons never branch back into biadjoint scalars.  Second, in YM theory the external sources generate both field strengths {\it and} gauge fields, so in the corresponding amplitudes in GBAS theory, the external legs can be either biadjoint scalars or gauge fields.  See  \Fig{fig:4pt} for the Feynman diagrams that contribute to three- and four-point scattering in this alternative formulation of YM theory.

Next, consider the {\it subset} of Feynman diagrams in which the gauge field does not appear, {\it i.e.}~diagrams involving field strengths only and no gauge fields.  This effectively replaces $D^2$ with $\Box$ in \Eq{EOM_YM}.  
Amusingly, these diagrams satisfy color-kinematics duality off-shell.  To understand why, let us compute the {\it off-shell} four-point subdiagram of field strengths currents embedded within a larger diagram.  The Feynman diagrams in the $s$, $t$, and $u$ channel take the form of \Eq{stu_diagrams} but with kinematic numerators given by
\eq{
n_s &=8\,  \Pi^{\mu_1 \nu_1 \alpha}_{\;\;\;\;\;\;\;\;\;\; \beta} \Pi^{\mu_2 \nu_2 \beta}_{\;\;\;\;\;\;\;\;\;\; \gamma} \Pi^{\mu_3 \nu_3 \gamma}_{\;\;\;\;\;\;\;\;\;\; \delta} \Pi^{\mu_4 \nu_4 \delta}_{\;\;\;\;\;\;\;\;\;\; \alpha} - \{ 3 \leftrightarrow 4\} \\
n_t &= 8\,  \Pi^{\mu_2 \nu_2 \alpha}_{\;\;\;\;\;\;\;\;\;\; \beta} \Pi^{\mu_3 \nu_3 \beta}_{\;\;\;\;\;\;\;\;\;\; \gamma} \Pi^{\mu_1 \nu_1 \gamma}_{\;\;\;\;\;\;\;\;\;\; \delta} \Pi^{\mu_4 \nu_4 \delta}_{\;\;\;\;\;\;\;\;\;\; \alpha} - \{ 1 \leftrightarrow 4\} \\
n_u &= 8 \,  \Pi^{\mu_3 \nu_3 \alpha}_{\;\;\;\;\;\;\;\;\;\; \beta} \Pi^{\mu_1 \nu_1 \beta}_{\;\;\;\;\;\;\;\;\;\; \gamma} \Pi^{\mu_2 \nu_2 \gamma}_{\;\;\;\;\;\;\;\;\;\; \delta} \Pi^{\mu_4 \nu_4 \delta}_{\;\;\;\;\;\;\;\;\;\; \alpha} - \{ 2 \leftrightarrow 4\}   .
}{}
The numerators are trivially related by relabeling and sum to
\eq{
n_s + n_t+n_u &= 0 ,
}{}
so the kinematic Jacobi identities are satisfied automatically.  Said another way, the $F^3$ theory exhibits manifest color-kinematics duality as expected.

The amplitudes of $F^3$ theory are perfectly consistent if the external polarizations are interpreted as generic antisymmetric tensors.  However, when these antisymmetric tensors are equated with field strengths, the resulting amplitudes are no longer gauge invariant.  To restore gauge invariance one must of course include all interactions involving both the field strengths and the gauge fields.  Formally, this is equivalent to a replacement of $\Box$ with $D^2$.   This is precisely what is meant when we say that the formulation of YM theory in \Eq{EOM_YM} exhibits covariant color-kinematics duality.  

\subsubsection{EYM $\otimes$ F${}^3$ = GR}

Covariant color-kinematics duality implies that GR is the covariant double copy of EYM theory and $F^3$ theory. \Eq{replace_YM} maps the gauge field and field strength of EYM to
\eq{
A^{a}_\mu \quad &\arrowYM\quad \omega_{\mu\ov \mu \ov \nu} \\
F^{a}_{\mu\nu} \quad &\arrowYM\quad R_{\mu \nu \ov \mu \ov \nu} ,
}{AF_to_wR}
which we recognize immediately as the spin connection and Riemann curvature tensor.  The barred and unbarred indices should be interpreted as tetrad and metric indices, respectively, and can in principle be independent.  However,  since we are interested in GR here, {\it i.e.}~gravity sans dilaton and two-form, we will eventually impose the usual symmetry of the Riemann tensor on the exchange of the barred and unbarred indices.

As a sanity check we apply \Eq{replace_YM} to the relation between the gauge field and field strength, 
\eq{
F^a_{\mu\nu} = \partial_{[\mu} A^a_{\nu]} + f^{abc} A^b_\mu A^c_\nu \quad\arrowYM \quad R_{\mu\nu\ov\mu \ov \nu} = \partial_{[\mu} \omega_{\nu] \ov \mu \ov\nu} + 
\omega_{\mu \ov \rho \ov  \mu}  \omega^{\;\; \;\; \ov\rho}_{\nu \ov \nu} -  \omega_{\nu \ov \rho \ov  \mu}  \omega^{\;\; \;\; \ov\rho}_{\mu \ov \nu},
 }{}
 thus deriving the familiar relation between the spin connection and the Riemann tensor. 
  Last but not least, we find that \Eq{replace_YM} maps the EYM equations of motion to
\eq{
\nabla^\mu F_{\mu\nu}^a + f^{abc}  A^{b \mu} F_{\mu\nu}^c= J^{a}_\nu \quad \arrowYM \quad \nabla'^\mu R_{\mu\nu \ov \mu\ov \nu} + \omega^{ \mu}_{\;\; \ov \rho \ov \mu} R_{\mu\nu \ov \nu}^{\;\;\;\;\;\;\ov \rho} - \omega^{ \mu \;\; \ov\rho}_{\;\; \ov \nu} R_{\mu\nu\ov \rho \ov \mu}= \nabla_{[\ov \mu} T_{  \ov \nu] \nu} ,
 }{EYM_to_GR}
 where the color current is mapped to the derivative of the energy-momentum tensor.   
 
 Note that the derivative $\nabla$ appearing in \Eq{EYM_to_GR} is a gravitational covariant derivative which depends implicitly on the graviton through the Christoffel symbol. Furthemore, we have defined $\nabla'$ as a gravitational covariant derivative which acts on the Riemann tensor as if its barred spacetime indices were absent.  This object only appears because our replacement rules simply append indices onto the field strength in order to generate the Riemann tensor.  

As one might expect, the $\nabla'$ and spin connection terms in \Eq{EYM_to_GR} combine to form a bona fide $\nabla$ with appropriate connection terms for all the indices of the Riemann tensor.  Contracting \Eq{EYM_to_GR} into  
the tetrads $e^{\ov \mu}_\rho e^{\ov\nu}_\sigma$, we obtain
\eq{
 \nabla^\mu R_{\mu\nu \rho\sigma} = \nabla_{[\rho} T_{  \sigma]\nu}.
}{EinsteinEq}
Now {\it imposing} that the Riemann tensor is symmetric under swapping its $\mu\nu$ and $\rho\sigma$ indices, we find that \Eq{EinsteinEq} is equivalent via the second Bianchi identity to the derivative of Einstein's equations in natural units where $8\pi G=1$.    Hence, GR is equivalent to EYM theory with the gauge field and field strength dressed with additional spacetime indices.\footnote{It is also possible to derive GR directly from a theory of biadjoint scalars that are minimally coupled to gravity and doubly gauged with respect to color and dual color.  Here we apply the $\otimes \; F^3$ replacement rules to both colors.  Equating the spin and dual spin connections and projecting out trace modes, we obtain an equation for the Riemann tensor of the form $\nabla^2 R + R^2 \sim \nabla^2 T$, which is by construction isomorphic to the equation of motion for the field strength, $D^2 F + F^2 \sim D J$.  Amusingly, this covariant double copy formulation of GR is literally ``textbook'': it is the de Rham wave equation (see exercise 15.2 of Misner, Thorne, and Wheeler \cite{MTW}). }  So in the context of gravity, covariant color-kinematics duality is a trivial manifestation of the tetrad formalism.

\subsubsection{Classical Double Copy}

The entirety of our analysis holds at the level of equations of motion.  So it is natural to ponder its implications for the classical double copy \cite{ClassicalDC1, ClassicalDC2,ClassicalDC3,ClassicalDC4,ClassicalDC5, ClassicalDC6,ClassicalDC7,ClassicalDC8,ClassicalDCTwistor1,ClassicalDCTwistor2}.  Amusingly, covariant color-kinematics duality enforces a trivial version of the classical double copy which applies to {\it any vacuum solution} of Einstein's equations and is covertly a restatement of the tetrad formalism.  The claim is as follows.  For any vacuum solution of GR, compute the corresponding spin connection and Riemann curvature tensor.   Now simply {\it define} these objects to be the gauge field and field strength of some exotic gauge theory, as shown in the mapping in \Eq{AF_to_wR}.  Each antisymmetric pair of spacetime indices should be interpreted as an internal color index.  Then \Eq{EYM_to_GR} guarantees that these fields will automatically satisfy the equations of motion for YM theory propagating in a background spacetime defined by the original GR solution.  Bear in mind, this variant of the classical double copy applies to solutions in which color plays a critical role.  

For this to work one must actively ignore the back-reaction of the YM dynamics on the underlying spacetime metric.  Diagrammatically, these effects correspond to scattering processes in which gravitons are emitted and then reabsorbed by the YM field.  These are exactly analogous to the amplitudes of GBAS theory in which gauge fields are emitted and then reabsorbed by the biadjoint scalar.  As described earlier, such contributions go beyond the single trace contributions for which covariant color-kinematics duality applies.   Thus, spacetime curvature dynamically evolves precisely as if it were a ``probe'' gauge field propagating in that very same background spacetime.   Note that an analogous procedure also maps solutions of pure YM theory onto those of GBAS theory in a color background.

Even though any solution of GR can be recast as a solution of YM theory in curved spacetime, the converse is {\it not true}, at least in general.  Indeed, the opposite procedure only applies to a very special class of solutions of YM theory in which the gauge field coincides exactly with the spin connection of the background.   Consequently, this construction is as general as it is useless for  building new solutions of GR.

Nevertheless, it is entertaining to study this version of the classical double copy in its most basic incarnation: a four-dimensional, Euclidean black hole.  
In order to map the spin connection $\omega_{\mu  \ov \mu\ov\nu}$ to the gauge field $A_{\mu}^a$, we simply repackage the antisymmetric pair of four-dimensional spacetime indices $\ov \mu\ov\nu$ as a single internal six-dimensional color index $a$.  Concretely, we define $
\omega_{\mu  \ov \mu\ov\nu} = A_{\mu}^a \lambda^a_{\ov \mu \ov \nu}$, where the antisymmetric matrices $\lambda^a_{\ov \mu \ov\nu}$ parameterize the six generators of four-dimensional Euclidean rotations.  
In Euclidean Schwarzschild coordinates, the gauge field configuration corresponding to a black hole is
\eq{
A_\mu^a \sim 
\begin{pNiceMatrix}[columns-width=auto]
-\frac{R_S}{2r^2} 		& 	0				& 	0						 & \phantom{\sqrt{\frac{A}{B}}} 0 \phantom{\sqrt{\frac{A}{B}}}  	  &  \cdots   \\
0				& 	0				& 	\phantom{\sqrt{\frac{A}{B}}} 0 \phantom{\sqrt{\frac{A}{B}}} 			 &  0  &\cdots\\
0				& 	\sqrt{1-\frac{R_S}{r}}	&     0	 			&   0 &\cdots\\
 0 					& 	0				& 	\sqrt{1-\frac{R_S}{r}} \, \sin\theta						 &  \cos\theta &\cdots 
\end{pNiceMatrix},
}{A_DC}
where the rows and columns correspond to the $\mu$ and $a$ indices, respectively, and $R_S$ is the Schwarzschild radius.   Here we haven chosen a convenient basis for the generators $\lambda^a_{\ov \mu \ov\nu}$ in which the vanishing components of the gauge field are the rightmost entries of \Eq{A_DC} denoted by ellipses.  It is straightforward to verify that \Eq{A_DC} indeed satisfies the equations of motion of YM theory in a Euclidean Schwarzschild background, as expected.  Furthermore, it is easy to see that the associated color electric and magnetic fields scale as $\sim R_S / r^3$  at long distances.  

To summarize, we have shown that the spin connection of a Euclidean black hole behaves like a dipolar, probe gauge field configuration stabilized by the background spacetime curvature.  From this perspective, the Schwarzschild solution of GR has an alternative interpretation as a ``chromo-gravitational atom'' in which the roles of the proton and electron are played by the background metric and gauge field, respectively.
 
\subsection{Kinematic Current}

The framework we have just described implements the covariant double copy at the level of fields.  By applying the $\otimes \; F^3$ replacement rule directly to the dual color current of GBAS theory in \Eq{JK_BAS}, we can elegantly derive the corresponding covariant kinematic current of YM theory,
\eq{
\hlLOW{
{\cal K}^{\YM}_{\mu\nu\alpha} =  F^{a}_{\rho \mu} \overset{\leftrightarrow}{D}_\alpha F^{a\rho}_{\nu }.
}
}{K_YM}
This object is necessarily conserved by virtue of the fact that the kinematic structure constants of $F^3$ theory automatically satisfy the kinematic Jacobi identities.  We can verify this assertion via explicit calculation, where we find that
\eq{
\partial^\alpha {\cal K}^{\YM}_{\mu\nu  \alpha} =  F^{a}_{\rho \mu} \overset{\leftrightarrow}{D^2} F^{a\rho}_{\nu } = -f^{abc} ( F^{a\rho}_{\;\;\;\; \mu}F^{b}_{\sigma [\nu}  F^{c\sigma}_{\rho]}  - F^{a\rho}_{\nu }F^{b}_{\sigma [\rho}  F^{c\sigma}_{\mu]}) =0,
}{conserved_YM}
on the support of the YM equations of motion \Eq{EOM_YM} and in the absence of external sources.  

Just like for the NLSM, the covariant kinematic current of YM theory is secretly related to the energy-momentum tensor, which for YM theory is
\eq{
T_{\mu\nu}^{\YM} &= F_{\rho\mu}^a F_{\nu}^{a \rho} +\tfrac{1}{4} \eta_{\mu\nu} F^a_{\rho\sigma} F^{a\rho\sigma}.
}{}
After a bit of algebra, we find that \Eq{K_YM} is
\eq{
{\cal K}^{\YM}_{\mu\nu\alpha} & = -\partial_{[\mu} T_{\nu]\alpha}^{\YM} +D^\rho F_{\rho\alpha}^a F_{\mu\nu}^a - D^\rho F^a_{\rho [\mu} F^a_{\nu]\alpha} - \partial^\rho (F^a_{\rho\alpha} F^a_{\mu\nu} - F^a_{\rho [\mu} F^a_{\nu]\alpha}) -\tfrac{1}{4}\eta_{\alpha [\mu} \partial_{\nu]} (F_{\rho\sigma}^a F^{a\rho\sigma}) \\
& = -\partial_{[\mu} T_{\nu]\alpha}^{\YM} + \textrm{improvement terms}.
}{K_is_T}
In the second line we have dropped all terms proportional to the YM equations of motion and set aside improvement terms that are automatically conserved by antisymmetry.  \Eq{K_is_T} implies that the covariant kinematic current of YM theory is the derivative of the energy-momentum tensor.  As discussed earlier, any current which is itself the derivative of another current corresponds to a trivial charge.  Hence, the covariant kinematic current for YM theory cannot arise from a traditional symmetry in the usual sense.

\section{Applications}

\label{sec:applications}

We have shown from first principles how YM theory is the covariant double copy of GBAS theory and $F^3$ theory.   In this section we utilize this insight to derive a novel decomposition of YM amplitudes of the schematic form,
\eq{
A^{\YM} \quad \sim\quad  \sum A^{\GBAS} \, F \cdots F,
}{inv_trans_schematic}
{\it i.e.}~as a sum over GBAS amplitudes weighted by products of field strengths.  
Meanwhile, the inverse decomposition is encoded in the unifying relations \cite{transmutation}, which are of the form
\eq{
A^{\GBAS} \quad \sim \quad   {\cal T} \cdots {\cal T} A^{\YM}   .
}{trans_schematic}
Here the transmutation operators ${\cal T}$ are simply differentials acting on the space of kinematic invariants.    By applying ${\cal T}$ repeatedly to \Eq{inv_trans_schematic} and using \Eq{trans_schematic}, we can solve for all GBAS amplitudes purely in terms of BAS amplitudes.  Plugging back into \Eq{inv_trans_schematic}, we  obtain a formula for all YM amplitudes of the form
\eq{
A^{\YM} \quad \sim\quad  \sum A^{\BAS} \,  {\cal T} \cdots {\cal T} \, F \cdots F.
}{BCJ_schematic}
As we will see, the kinematic functions multiplying the BAS amplitudes in \Eq{BCJ_schematic} automatically satisfy the kinematic Jacobi identities, {\it i.e.}~they are BCJ numerators.  Conveniently, we are able to derive a compact, closed-form expression for all such BCJ numerators in YM theory, and similarly in the NLSM.   By the standard double copy prescription, these explicit formulas generate closed-form expressions for all amplitudes in YM, GR, NLSM, SG, and BI.

Another avatar of color-kinematics duality is the fundamental BCJ relation, which reduces the naive number of independent color-ordered amplitudes \cite{BCJ1}.  Amusingly, this relation is rather straightforwardly derived from equations of motion, which we also describe in \App{app:BCJ}.

\subsection{Color Structures}

Earlier we described how the $\otimes \, F^3$ replacement rules directly map color structure constants onto kinematic structure constants.  Let us now apply this substitution to arbitrary color topologies.  In particular, \Eq{replace_YM} sends a generic color half ladder \cite{DDM} to
\eq{
f^{a_1 a_2 a_3} &\quad  \arrowYM \quad  ( -i  ) \,  {\rm tr} [[F_1,F_2]  \widetilde F_3]\\
f^{a_1 a_2 a_3 a_4} &\quad   \arrowYM \quad   ( - i )^{2} \,  {\rm tr} [[[F_1,F_2],F_3]  \widetilde F_4]\\
& \;\; \quad \vdots \\
f^{a_1 a_2 a_3 \cdots a_n} &\quad   \arrowYM \quad   ( -i )^{n-2} \,  {\rm tr} [[\cdots [[F_1,F_2],F_3],  \cdots , F_{n-1} ] \widetilde F_n],
}{f_to_F}
where $f^{a_1 a_2 a_3 \cdots a_n} = f^{a_1 a_2 b_1}f^{b_1 a_3 b_2}  \cdots f^{b_{n-3} a_{n-1} a_n}$ and we have by convention designated the last leg in each half ladder to be the root leg. 
Here we have also defined linearized field strength tensors for each leg,
\eq{
\left[F_i\right]_{\mu\nu} &= p_{i \mu} \varepsilon_{i \nu} - p_{i \mu} \varepsilon_{i \nu} \qquad \textrm{for} \qquad i\neq n   \\
{}[ \widetilde F_n  ]_{\mu\nu}  &= \frac{1}{p_n q} \left(q_{\mu} \varepsilon_{n \nu} - q_{ \mu} \varepsilon_{n \nu} \right) ,
}{Fdef1}
as well as a trace over {\it spacetime} indices
\eq{
{\rm tr}\left({\cal O} \right) = \tfrac{1}{2}{\cal O}_{\mu\nu} \eta^{\mu\nu}.
}{Odef}
The nested commutators in \Eq{f_to_F} implement the Feynman vertex in \Eq{vertex_F3}.  Meanwhile, the field strengths $F_i$ are the leaf leg polarizations in \Eq{leaf_pol_F3} and the field strength $\widetilde F_n$ is the root leg polarization in \Eq{root_pol_F3}.  

As per the usual relations governing color structure constants and generators in \Eq{normalizations},  \Eq{f_to_F} implies a mapping between trace color structures and  field strengths,
\eq{
{\rm tr}\left[T^{a_1} T^{a_2} T^{a_3} \right] & \quad  \arrowYM \quad    {\rm tr} [F_1 F_2  \widetilde F_3]\\
{\rm tr}\left[T^{a_1} T^{a_2} T^{a_3} T^{a_4} \right] & \quad  \arrowYM \quad   {\rm tr} [F_1 F_2 F_3  \widetilde F_4]\\
& \;\; \quad \vdots \\
{\rm tr}\left[T^{a_1} T^{a_2}  T^{a_3}  \cdots  T^{a_{n-1}} T^{a_n} \right]  &\quad   \arrowYM \quad   {\rm tr} [F_1 F_2 F_3 \cdots F_{n-1} \widetilde F_n],
}{T_to_F}
so the field strength tensors are themselves generators of the kinematic algebra in the fundamental representation.  

Since these field strength traces appear ubiquitously throughout our analysis, it will be worthwhile to define a bit of convenient shorthand notation.  For an arbitrary ordered set $\sigma = \sigma_1 \sigma_2 \sigma_3 \cdots \sigma_\ell$, we define the corresponding trace of field strengths,
\eq{
F[\sigma n]& = {\rm tr}[ F_{\sigma} \widetilde F_n ],
}{trFdef}
where the product of field strength tensors,
\eq{
F_\sigma = \prod_i^{|\sigma|} F_{\sigma_i},
}{Fdef2}
should be evaluated in the specific ordering $\sigma$, and where $|\sigma|=\ell$ is the size of the set.
Throughout our discussion, an ordered set $\sigma\tau$ will denote the concatenation of the ordered sets $\sigma$ and $\tau$, so for example $\sigma n =  \sigma_1 \sigma_2 \sigma_3 \cdots \sigma_\ell n$.

\subsection{Field Strength Decomposition}

We are now prepared to derive the field strength decomposition of YM amplitudes into GBAS amplitudes and field strengths.  First, recall the standard color decomposition of the GBAS amplitude (see \cite{ElvangHuang, DixonReview1, DixonReview2, TASIReview} for reviews),
\eq{
A^{\GBAS}_{\phi n}
= \sum_{\sigma\in S(\phi)
} A^{}[\sigma n] \, {\rm tr} \left[T^{a_{\sigma_1}}   \cdots  T^{a_{\sigma_\ell}} T^{a_n} \right].
}{A_GBAS_decomp}
Here $\phi n = \phi_1  \cdots \phi_\ell n $ denotes the set of all biadjoint scalars in the amplitude, which by convention will always include the root leg.  All gauged color structures, {\it i.e.}~the color indices shared by both the gauge fields and biadjoint scalars,  are implicit and so we should think of \Eq{A_GBAS_decomp} as color-stripped on those indices.   In contrast,  the explicit trace structures in \Eq{A_GBAS_decomp} correspond to the ungauged color index that is only carried by the biadjoint scalars.

From \Eq{EOM_YM} we saw that the equation of motion for the field strength of YM theory is the same as that of GBAS theory in \Eq{EOM_GBAS} upon application of the $\otimes \, F^3$ replacement rule.  At the same time, the field strength in \Eq{EOM_YM} is sourced by the derivative of the same external current that sources the gauge field.
Thus, the $\otimes \, F^3$ replacement rule sends
\eq{
\sum_{\phi \in {\mathbb P}^+(1\cdots n-1)} A^{\GBAS}_{\phi n} \quad \arrowYM\quad
 \sum_{\phi \in {\mathbb P}^+(1\cdots n-1)} \sum_{\sigma \in S(\phi)} A^{}[\sigma   n] \, F[\sigma n] .
}{AGBAS_to_AYM}
Here the left-hand side is the sum over all GBAS amplitudes with the root leg taken to be a scalar and all leaf legs taken to be either a gauge field {\it or} a scalar.   Recall that we saw this earlier in the Feynman diagrams for YM theory shown in \Fig{fig:4pt}.
  The first sum in \Eq{AGBAS_to_AYM} runs over ${\mathbb P}^+(x)$, which is the nonempty power set, {\it i.e.}~set of all nonempty subsets, of the set $x$.  The second sum runs over permutations $S(x)$ of the set $x$.

The left-hand side of \Eq{AGBAS_to_AYM} would ordinarily be a nonsensical quantity, since it is a sum of amplitudes with different numbers of gauge fields and thus different little group weights.  But crucially, the kinematic structures introduced by the $\otimes \, F^3$ replacement rule carry the exactly appropriate little group weight to cancel this.  Consequently, the right-hand side of \Eq{AGBAS_to_AYM} is perfectly well-defined and homogenous under little group scaling.  In fact, as a consequence of covariant color-kinematics duality, it is precisely an expression for the YM amplitude,
\eq{
\hlHIGH{
A_n^{\YM} = \sum_{\phi \in {\mathbb P}^+(1\cdots n-1)} \sum_{\sigma \in S(\phi)} A^{}[\sigma   n] \, F[\sigma n],
}
}{field_strength_decomp}
decomposed into GBAS amplitudes times products of field strengths.  This formula is reminiscent of a similar decomposition discovered in \cite{FeiTeng, BoFeng}, which marked two rather than one leg as special and involved a different function of the field strengths.
Evaluated for the three-point YM amplitude, \Eq{field_strength_decomp} becomes
\eq{
A_3^{\YM} &= A^{}[123] \, F[123] + A^{}[213]\,  F[213] +A^{}[13] \, F[13]+ A^{}[23] \, F[23],
}{}
while the four-point YM amplitude is
\eq{
A_4^{\YM} =& \phantom{{}+{}}
A[1234] \, F[1234]
+A[1324]\, F[1324]
 +A[2134] \,F[2134]\\
& +A[2314] \,F[2314]+A[3124] \,F[3124]
 +A[3214] \, F[3214]\\
& + A[124] \,F[124]+A[214] \, F[214]
 +A[134]  \,F[134]\\
& +A[314] \, F[314]+A[234] \,F[234] +A[324]  \,F[324] \\
 & +A[14] \, F[14]+A[24] \, F[24]
 +A[34] \,F[34],
 }{}
 and so on for five-point scattering and higher.    We have verified the validity of \Eq{field_strength_decomp} up to seven-point scattering in YM theory.

Last but not least, one can doubly apply the $\otimes \, F^3$ replacement rule to derive a similar field strength decomposition for gravitational amplitudes,
\eq{
\hlHIGH{
A_n^{\GR} = \sum_{\substack{\phi \in {\mathbb P}^+(1\cdots n-1) \\ \bar \phi \in {\mathbb P}^+(1\cdots n-1)}} \sum_{\substack{\sigma \in S(\phi) \\ \bar \sigma \in S(\bar \phi)}} A^{}[\sigma   n| \bar\sigma n] \, F[\sigma n]  \, \bar F[\bar\sigma n],
}
}{field_strength_decomp_grav}
where the barred field strength trace depends on a new set of barred polarizations. On the right-hand side is the doubly ordered amplitude of biadjoint scalars, gauge fields, and the extended gravity multiplet, {\it i.e.}~the graviton, dilaton, and two-form.   The intersection $\sigma \cap \bar\sigma$ denotes the scalars, the difference sets $\sigma - \bar \sigma$ and $\bar\sigma-\sigma$ denote the gauge fields, and the remainder denotes the gravitational states.  We have checked via explicit calculation that \Eq{field_strength_decomp_grav} is valid up to six-point scattering in gravity.   

The field strength decompositions for YM theory and gravity in \Eqs{field_strength_decomp,field_strength_decomp_grav} are some of the main results of this paper. These equations reflect the intriguing physical picture encountered earlier: field strengths evolve like charged, self-interacting scalars and curvatures evolve like gravitating, self-interacting field strengths. 

\subsection{Inverse Transmutation}

As discussed in \cite{transmutation}, transmutation is a differential operation that maps the amplitudes of YM theory onto those of GBAS theory,
\eq{
A[123\cdots \ell  n] &=  {\cal T}[123\cdots \ell n]  \, A_n^{\YM}.
}{transmute1}
Here the transmutation operator is a product of sequential operations,
\eq{
 {\cal T}[123\cdots \ell n]  =  {\cal T}_{1\, n} \prod_{i=1}^{\ell-1} {\cal T}_{i\,  i+1 \, n} ,
 }{transmute2}
which are differentials in the space of kinematic invariants,
\eq{
{\cal T}_{i\, j} = \frac{\partial}{\partial (\varepsilon_i \varepsilon_j)} \qquad \textrm{and} \qquad   {\cal T}_{i \, j \, k} =
\frac{\partial}{\partial (p_k \varepsilon_j)} - \frac{\partial}{\partial (p_i \varepsilon_j)} .
}{}
As is obvious from the basic structure of \Eq{transmute1}, transmutation is a tool for building amplitudes with {\it fewer} gauge fields from those with {\it more}.  However, it should also be self-evident that the reverse operation would be of far greater use for practical applications.  Here we use the field strength decomposition to derive precisely such a procedure, dubbed ``inverse transmutation'', which constructs amplitudes with {\it more} gauge fields from those with {\it fewer}.   

Inserting the field strength decomposition of YM theory in \Eq{field_strength_decomp} directly into \Eq{transmute1}, we obtain the lengthy expression,
\eq{
A[123\cdots \ell n] &= \sum_{\theta \in {\mathbb P}(\ell+1 \cdots n-1)} \sum_{\tau \in S(\theta)} \sum_{i=1}^{\ell}  \sum_{\rho \in \tau \shuffle (i+1\cdots \ell)   }A[1\cdots i \rho n]  \, \xi(i,\tau) \, 
 {\cal T}[1\cdots i  n] \, F[1\cdots i \tau n]    \\
&=\sum_{\theta \in {\mathbb P}(\ell+1 \cdots n-1)} \sum_{\tau \in S(\theta)} \sum_{i=1}^{\ell}  \sum_{\rho \in \tau \shuffle (i+1\cdots \ell)   }A[1\cdots i \rho n] \left\{ -  \frac{p_i F_{\tau}  q}{p_n q} \right\},
}{spin_reduce1}
where we have defined the parameter
\eq{
\xi(i,\tau) &= \left\{
\begin{array}{ll}
1& ,\quad  i=1 \textrm{ and } |\tau| = 0 \\ 
2& , \quad \textrm{else} \\ 
\end{array} \right. ,
}{}
and ${\mathbb P}(x)$ is the power set, {\it i.e.}~the set of all subsets including the empty set, of the set $x$.  Here $ x \shuffle y$ denotes the shuffle product of $x$ and $y$, which is the set of all sets in which $x$ and $y$ are merged so as to maintain the relative order of all elements within $x$ and $y$ separately.   These shuffle products appear because the transmutation operator  $  {\cal T}_{i \, i+1 \, n}$ generates the sum of amplitudes in which scalar leg $i+1$ is inserted between scalar legs $i$ and $n$ in all possible positions amongst the scalar legs already residing between $i$ and $n$.

To derive the first line of \Eq{spin_reduce1} we have used the fact that a sequence of transmutation operators of the form ${\cal T}_{i \, i+1\, n}$ will annihilate the field strength trace unless the sequence is a contiguous set of adjacent legs.  This happens because ${\cal T}_{i \, i+1\, n}$ extracts the kinematic invariants $p_i \varepsilon_{i+1}$ and $p_n \varepsilon_{i+1}$, which only appear in the field strength trace if $F_{i+1}$ appears either adjacent to $F_i$ or $\widetilde F_n$.
The second line of  \Eq{spin_reduce1} is then obtained by plugging in the explicit formulas for the transmutation operators and the field strength trace.

Now we observe that $A[123\cdots \ell n]$ appears on both the left- and right-hand sides of \Eq{spin_reduce1}, appearing in the latter through the summand in which $\theta$ is the empty set.  Thus we can shuffle all dependence on $A[123\cdots \ell n]$ to the left-hand side of the equation and solve for it, yielding
\eq{
A[123\cdots \ell n] &= \sum_{\theta \in {\mathbb P}^+(\ell+1\cdots  n-1)} \sum_{\tau \in S(\theta)} \sum_{i=1}^{\ell}  \sum_{\rho \in  \tau \shuffle (i+1 \cdots\ell)  }A[1\cdots i \rho n] \left\{ -    \frac{p_i F_{\tau}  q}{(p_1 +\cdots + p_{\ell} + p_n) q} \right\}.
}{spin_reduce2}
Since the sum now runs over the nonempty power set, the right-hand side only involves amplitudes with strictly fewer gauge fields than the right-hand side.  Note also that \Eq{spin_reduce2} holds only if $\ell < n-1$.  If $\ell=n-1$, then the left- and right-hand sides of \Eq{spin_reduce1} are trivially equal to the same pure BAS amplitude. In this case  \Eq{spin_reduce1} is automatic and one cannot solve for the left-hand side, which is why \Eq{spin_reduce2} has a division by zero when $\ell=n-1$.

Next, let us generalize \Eq{spin_reduce2} to an arbitrary ordering of scalars $\sigma n =\sigma_1 \sigma_2 \sigma_3 \cdots \sigma_\ell n$, yielding an inverse transmutation relation
\eq{
A[\sigma n] &= \sum_{\theta \in {\mathbb P}^+(\overline{ \sigma n})} \sum_{\tau \in S(\theta)} \sum_{i=1}^{|\sigma|}  \sum_{\rho \in   \tau \shuffle \sigma_{>i} }A[\sigma_{\leq i} \rho n] \left\{ - \frac{p_{\sigma_i} F_{\tau}  q_\sigma}{p_{\sigma n} q_\sigma} \right\},
}{spin_reduce3}
which decomposes an amplitude with more gauge fields into those with fewer.   In \Eq{spin_reduce3} we have defined $\overline{\sigma n}$ to be the complement of $\sigma n$, and 
\eq{
p_\sigma = \sum_i^{|\sigma|} p_i,
}{p_sigma}
to be the total momentum flowing through any set of external legs $\sigma$.  Meanwhile, $\sigma_{< i}$ is the set of elements in $\sigma$ to the left of $\sigma_i$, and  $\sigma_{>i}$ is the set of elements in $\sigma$ to the right of $\sigma_i$.   We similarly define $\sigma_{\leq i}$ and $\sigma_{\geq i}$ as the corresponding inclusive sets.   In this notation the full ordered set is $\sigma = \sigma_{<i} \sigma_i \sigma_{>i}$.  As before, \Eq{spin_reduce3} applies only if $|\sigma| < n-1$, since the degenerate case of $|\sigma|=n-1$ reduces to the pure BAS amplitude.

As written, \Eq{spin_reduce2} depends on a single reference $q$, but we are actually permitted to use a {\it different} reference for each possible GBAS amplitude.  Consequently, for the general formula in \Eq{spin_reduce3} we have lifted the single reference $q$ to a distinct reference $q_\sigma$ specific to {\it each} ordering $\sigma$.  This immense freedom will come in handy later on.

For reasons that will soon become clear, let us now reorganize the terms in \Eq{spin_reduce3} according to whether they are ``in order'' or ``out of order'' with respect to a canonical ordering of our choice.  For concreteness, we take this canonical ordering to be the literal numerical ordering of the natural numbers.  Thus, we are interested in separating out those terms in \Eq{spin_reduce3} for which the legs in $A[\sigma_{\leq i} \rho n]$ are numerically ordered.    This is only possible if all the elements of $\rho$ are numerically greater than all the elements of $\sigma_{\leq i}$.  Since $\tau$ is a subset of $\rho$ this implies that $\tau$ is numerically greater than all the elements of $\sigma_{\leq i}$.  Regrouping terms in \Eq{spin_reduce3} then yields
\eq{
A[\sigma n] &= \sum_{\tau \in {\mathbb P}^+(\overline{ \sigma n})} A[\textrm{sort}(\sigma,  \tau) n]  \; G[\sigma_{<\tau}, \tau, \sigma_{>\tau} n]
+ \textrm{out of order},
}{spin_reduction_order}
where ${\rm sort}(\sigma,\tau)$ is the union of $\sigma$ and $\tau$ sorted in numerical order, $\sigma_{<\tau} $ is the set of elements in $\sigma$ which are numerically less than the minimal element of $\tau$, and $\sigma_{>\tau} $ is the set of elements in $\sigma$ which are numerically greater than the minimal element of $\tau$.  The terms not shown explicitly all involve amplitudes whose legs are not in numerical order.
In \Eq{spin_reduction_order} we have introduced the kinematic function
\eq{
G[\sigma, \tau, \rho n]  &= -  \frac{p_{\sigma} F_{\tau}  q_{\sigma\rho}}{p_{\sigma\rho n} q_{\sigma\rho}} ,
}{Gdef}
where $F_\tau$ is defined as in \Eq{Fdef2}, $p_\sigma$ and $p_{\sigma \rho n}$ are defined as in \Eq{p_sigma}, and $q_{\sigma\rho}$ is the reference momentum for that particular ordering.  Throughout, we either assume that $q_{\sigma\rho}$ is constant, {\it i.e.}~identical for all orderings, or that it depends only on the momenta in the set $\sigma\rho$.  Either choice ensures that permuting the ordering in the subscript argument of $q_{\sigma\rho}$ is equivalent to permuting the kinematic variables within.

\subsection{Analytic Formulas for Amplitudes}

There has been considerable effort dedicated to the explicit evaluation of BCJ numerators \cite{LionelNumerators,EdisonNumerators, SongHeNumerators, HenrickNMHVNumerators,CongkaoHQET, HeBCJNum, MafraBCJNum, FeiTeng, HenrikSDYM, ChenBCJNum, DuBCJNum, FengBCJNum1, FengBCJNum2, FuBCJNum, Bjerrum-BohrNumerators1,Bjerrum-BohrNumerators2, MafraNumerators1, MafraNumerators2, JJVirtuous, NaculichVirtuous}.   However, a completely general, closed-form, analytic expression has proven elusive.\footnote{Of course, it has long been known how to obtain BCJ numerators from known YM amplitudes taken as {\it input} \cite{Bjerrum-Bohr:2010pnr}.   } In this section, we show how our field strength decomposition, together with inverse transmutation, enables an analytic derivation of all BCJ numerators in YM theory and the NLSM.  Our resulting expressions constitute closed formulas for the scattering amplitudes for these theories for any number of external legs and in arbitrary dimensions.\footnote{While there exist several formulas for all tree-level YM amplitudes in the literature \cite{SYMFormula, YMFormula}, these all apply specifically to four dimensions and are expressed in terms of sums over paths.} Via the usual double copy procedure these also generate closed formulas for all scattering amplitudes in GR, SG, and BI theory.

\subsubsection{Yang-Mills Theory and Gravity}

By applying \Eq{spin_reduction_order} repeatedly on \Eq{field_strength_decomp}, one can reduce any GBAS amplitude into a basis of pure BAS amplitudes.  The resulting formula is
\eq{
\hlHIGH{
A_n^{\YM} = \sum_{\sigma \in S(1\cdots n-1)} A^{}[\sigma   n] \, K[\sigma n],
}
}{result1}
where  $A[\sigma n]$ are pure BAS amplitudes and $K[\sigma n]$ are the BCJ numerators of YM theory, expressed here in trace basis rather than the usual half ladder basis \cite{DDM}.  Crucially, $K[\sigma n]$ is permutation invariant on the $n-1$ legs in $\sigma$, {\it i.e.}~the BCJ numerator for general $\sigma$ is obtained by permuting the kinematic variables for a {\it single} ordering. Thus we need only define this expression for a single representative BCJ numerator, which we choose to be in numerical ordering, 
\eq{
\hl{
K[123\cdots n] = \sum_{\tau \in \textrm{part}(1\cdots n-1)} F[\tau_1 n]\prod_{i=2}^{|\tau|} G[ ( \tau_1 \cdots \tau_{i-1})_{<\tau_i},\tau_i,(\tau_1 \cdots \tau_{i-1})_{>\tau_i} n].
}
}{result2}
For convenience, let us again write the explicit formulas for the trace of field strengths and the kinematic function derived previously in \Eqs{trFdef,Gdef}, 
\eq{
F[\sigma n]  = {\rm tr} [F_\sigma \widetilde F_n]  \qquad \textrm{and} \qquad
G[\sigma, \tau,\rho n]  = -  \frac{p_{\sigma} F_{\tau}  q_{\sigma\rho}}{p_{\sigma\rho n} q_{\sigma\rho}} ,
}{FandG}
where all subscripts either denote a set or concatenations of sets.  We defined earlier the trace in \Eq{Odef}, the fields strengths in \Eqs{Fdef1,Fdef2}, and the momentum of a set in \Eq{p_sigma}.

In \Eq{result2} we have introduced $\textrm{part}(x)$, which is the set of all ordered partitions of the set $x$ into subsets whose elements are in numerical order.   We also require that the first subset of every partition contains leg 1.   For example at low orders, $\textrm{part}(x)$ is
\eq{
{\rm part}(12) &=  \{ 12 \}, \{ 1,2 \}  \\
{\rm part}(123) &=  \{ 123 \}, \{ 12,3 \}, \{ 13,2 \} , \{ 1,23 \} , \{ 1,2,3 \} , \{ 1,3,2 \} \\
 {\rm part}(1234) &= \{1234\},\{134,2\},\{124,3\},\{123,4\},\{14,3,2\},\{14,2,3\},\{14,23\},\{13,4,2\}, \\
 & \phantom{{}={}}  \{13,2,4\},\{13,24\},\{12,4,3\},\{12,3,4\},\{12,34\},\{1,4,3,2\},\{1,4,2,3\}, \\
 & \phantom{{}={}} \{1,3,4,2\},\{1,3,2,4\},\{1,2,4,3\},\{1,2,3,4\},\{1,34,2\},\{1,24,3\},\{1,23,4\}, \\
 & \phantom{{}={}} \{1,4,23\},\{1,3,24\},\{1,2,34\},\{1,234\} ,
}{part_list}
and so on and so forth. 

The argument of the kinematic function in \Eq{result2} depends on $( \tau_1 \cdots \tau_{i-1})_{<\tau_i}$, which is the set of elements in $ \tau_1 \cdots \tau_{i-1}$ which are numerically less that the minimal element of  $\tau_i$, and $( \tau_1 \cdots \tau_{i-1})_{>\tau_i}$, which is the set of elements in $ \tau_1 \cdots \tau_{i-1}$ which are numerically greater that the minimal element of  $\tau_i$.  We emphasize here that numerical ordering is the relevant one for \Eq{result2} simply because our representative BCJ numerator has been chosen to be in numerical ordering.  
For the case of three-, four-, and five-point scattering, this numerator is
\eq{
K[123] =&  \phantom{{}+{}} F[123] + F[13]\,G[1,2,3] \\
K[1234] =&  \phantom{{}+{}} F[1234] +  F[124] \,G[12,3,4]+  F[134] \,G[1,2,34] \\
&+ F[14] \,G[1,23,4] +  F[14] \,G[1,2,4]\,G[12,3,4]+F[14]\, G[1,3,4] \,G[1,2,34]\\
K[12345] =& \phantom{{}+{}} F[12345]+F[1345]\,G[1,2,345]+F[1245]\,G[12,3,45]+F[1235]\,G[123,4,5] \\
&+F[145]\,G[1,2,345]\,G[1,3,45]+F[145]\,G[1,2,45]\,G[12,3,45]+F[145]\,G[1,23,45]\\
&+F[135]\,G[1,2,345]\,G[13,4,5]+F[135]\,G[1,2,35]\,G[123,4,5]+F[135]\,G[1,24,35]\\
&+F[125]\,G[12,3,45]\,G[12,4,5]+F[125]\,G[12,3,5]\,G[123,4,5]+F[125]\,G[12,34,5]\\
&+F[15]\,G[1,2,345]\,G[1,3,45]\,G[1,4,5]+F[15]\,G[1,2,45]\,G[1,4,5]\,G[12,3,45]\\
&+F[15]\,G[1,2,345]\,G[1,3,5]\,G[13,4,5]+F[15]\,G[1,2,35]\,G[1,3,5]\,G[123,4,5]\\
&+F[15]\,G[1,2,5]\,G[12,3,45]\,G[12,4,5]+F[15]\,G[1,2,5]\,G[12,3,5]\,G[123,4,5]\\
&+F[15]\,G[1,2,345]\,G[1,34,5]+F[15]\,G[1,24,5]\,G[12,3,45]\\
&+F[15]\,G[1,23,5]\,G[123,4,5]+F[15]\,G[1,4,5]\,G[1,23,45]\\
&+F[15]\,G[1,3,5]\,G[1,24,35]+F[15]\,G[1,2,5]\,G[12,34,5]+F[15]\,G[1,234,5] , 
}{}
where all terms are arranged in the same order as in \Eq{part_list}.
All other BCJ numerators can be obtained from the above expressions by literally permuting the arguments of the field strength traces and kinematic functions.  Since these objects only depend on the kinematic variables, one can also just permute the momenta and polarizations directly.

The BCJ numerators in \Eq{result2} automatically satisfy the kinematic Jacobi identities.  This is perhaps not so surprising given that these numerators were obtained by transmuting field strength traces which already manifestly exhibit the kinematic algebra and obey the kinematic Jacobi identities.  As a result, we can immediately insert our BCJ numerators into the usual double copy prescription to obtain
\eq{
\hlHIGH{
A_n^{\GR} = \sum_{\sigma \in S(1\cdots n-1)} \sum_{\bar\sigma \in S(1\cdots n-1)} A^{}[\sigma   n| \bar\sigma n] \, K[\sigma n]  \, \bar K[\bar\sigma n],
}
}{result3}
which is the gravitational scattering amplitude.
Here the barred BCJ numerator is also given by \Eq{result2}, but with all polarizations barred.

\Eqs{result1,result2,result3} are some of the principal results of this paper.  Altogether, they comprise explicit analytic expressions for all tree-level YM and gravity amplitudes for any number of external legs in arbitrary dimensions.  We have checked that these formulas correctly reproduce the amplitudes in YM theory up to seven-point scattering and in gravity up to six-point scattering.\footnote{The ancillary materials for this paper include \texttt{Mathematica} code computing all BCJ numerators assuming a universal choice for the reference momenta.  Code for computing all amplitudes in BAS theory can be found in \cite{MizeraPhiCubed} and the supplemental files therein.}

With respect to computational or algorithmic complexity, standard Feynman diagrams exhibit a hierarchy of the form GR $\gg$ YM $\gg$ BAS.  The double copy procedures of KLT or BCJ drastically alleviate this computational burden for gravity, sending GR $\sim$ YM $\gg$ BAS.   On the other hand, our formula for all BCJ numerators  in \Eq{result2} is a literal algebraic expression.  It does not secretly entail any recursive definitions, algorithmic procedures, sums over graphs, or unevaluated auxiliary integrals.  Thus, with the aid of \Eq{result2}, the amplitudes of YM and GR are no more complicated to calculate than those of BAS theory, thus placing GR $\sim$ YM $\sim$ BAS on equal footing.

The BCJ numerators in  \Eq{result2} have a number of advantageous properties.  First of all, for $n$-point scattering they manifest gauge invariance on $n-1$ leaf legs---though not for the root leg.  This happens because covariant color-kinematics duality preserves gauge invariance at every step.  At a technical level, this property arises because the BCJ numerators depend on the field strength trace $F[\sigma n]$ and kinematic function $G[\sigma,\tau,\rho n]$, which only depend on the polarizations of the $n-1$ leaf legs through field strengths.  As a result, the Ward identities are trivially satisfied on those legs.
Note that this does not contradict the results of \cite{NimaGaugeBootstrap}, since our BCJ numerators have spurious poles which depend on arbitrary reference momenta.   While the reference dependence {\it does not} cancel automatically within each BCJ numerator, it can still be chosen to eliminate all spurious poles, as we will see later on.

 Second, as previously noted, the BCJ numerators for $n$-point scattering are trivially related to each other by permuting the kinematic variables of the $n-1$ leaf legs.  Consequently, all BCJ numerators can be derived from the numerator for a single ordering, thus sidestepping the need to solve large systems of equations in order to enforce the kinematic Jacobi identities.

\subsubsection{Local Representation}

The BCJ numerators in \Eq{result2} have spurious poles on account of the denominator factors in the field strength trace $F[\sigma n]$ and kinematic function $G[\sigma,\tau,\rho n]$.   
We can, however, eliminate these nonlocalities by exploiting the  immense freedom afforded by the choice of reference momenta that appear in these quantities,  $q$ and $q_{\sigma\rho}$.   

We have not exhaustively explored the space of possible reference momenta.  Nevertheless, there are some obvious choices which substantially simplify our expressions for the BCJ numerators.
For example,  one option is to set $q=q_{\sigma \rho}$ together with
\eq{
p_n q = 1  \qquad \textrm{and} \qquad
p_i q = -\frac{1}{n-1} \qquad \textrm{for} \qquad i\neq n ,
}{smart_ref}
and $\varepsilon_i q =0$ for all $i$.  The numerical factors in \Eq{smart_ref} are fixed so as to maintain total momentum conservation.  Only the root leg is special, so the corresponding BCJ numerators are manifestly permutation invariant on the $n-1$ leaf legs.   By design, $F[\sigma n]$ and $G[\sigma,\tau,\rho n]$ are both local functions of the kinematic variables, albeit at the expense of manifest gauge invariance.  

Another interesting choice for the reference momenta is
\eq{
p_1 q  = -p_n q = 1 \qquad \textrm{and} \qquad
p_i q =  0 \qquad \textrm{for} \qquad i\neq 1,n,
}{}
with $\varepsilon_i q =0$ for all $i$ and $q_{\sigma\rho}$ unspecified.   In this case the resulting BCJ numerators only exhibit permutation invariance on $n-2$ legs.  Here the field strength traces take the particularly compact forms, $F[1\sigma n] =\tfrac{1}{2} \varepsilon_1 F_\sigma \varepsilon_n$ and $F[\sigma 1 n] =\tfrac{1}{2} \varepsilon_n F_\sigma \varepsilon_1$, and otherwise zero.  Half ladder objects of this type also appeared in the amplitudes decompositions of \cite{FeiTeng,BoFeng}.

\subsubsection{Spinor Helicity Representation}

Our expressions simplify further when we restrict to four dimensions.   To begin, let us consider a reference momentum vector $q$ which is null, so
\eq{
q = |q]\langle q| .
}{qF}
In principle, the reference momenta $q_{\sigma \rho}$ can be distinct for each ordering, but we instead set them all to equal to a single null momentum $r$, so
\eq{
 q_{\sigma\rho} = |r] \langle r|.
 }{}
Without loss of generality we also assume that the root leg has positive helicity and choose its polarization to depend on the same reference spinor $q$ as in \Eq{qF}, so\footnote{To simplify our expressions we normalize our polarization vectors without factors of $\sqrt{2}$, unlike those in \cite{Parke}.}
\eq{
\tilde  \varepsilon_n= \frac{|n] \langle q|}{\langle n q\rangle}.
}{}
Plugging this back into \Eq{Fdef1}, we obtain \eq{
F_{i^\pm} &=|i] [ i|  ,  |i\rangle \langle i| \qquad \textrm{for} \qquad i\neq n\\
 \widetilde F_n &= \frac{|q\rangle \langle q|}{\langle n q\rangle^2} ,
}{}
where the field strengths in the first line depend on the helicity of the corresponding leg.
For later convenience we also define the Parke-Taylor denominator factors,
\eq{
{\rm PT}(123\cdots \ell) = \langle 12\rangle \langle23 \rangle \cdots   \langle \ell1 \rangle \qquad \textrm{and} \qquad
\overline{\rm PT} ( 123\cdots \ell) = [ 12]  [23 ] \cdots  [\ell 1] .
}{}
Acting on a single element, these functions evaluate to ${\rm PT}(i) = \overline {\rm PT}(i) = 0$, while for the empty set  ${\rm PT}() = \overline {\rm PT}() = 2$, which is the trace of the identity matrix.  

Since the field strength trace and kinematic function in \Eq{FandG} are valid in any spacetime dimension, we can evaluate them in four dimensions to obtain
\eq{
F[\sigma n]  &\overset{\vsp d=4 \vsp}{\rightarrow} \frac{{\rm PT}(\sigma^- q) \overline{\rm PT}(\sigma^+)}{\langle q n\rangle^2}  \\
G[\sigma, \tau,\rho n] 
&\overset{\vsp d=4 \vsp}{\rightarrow} - \frac{1}{\langle r | p_{\sigma\rho n}|r]}\sum_{i=1}^{| \sigma|} 
\frac{{\rm PT}(\sigma_i \tau^- r) \overline{\rm PT}(\sigma_i \tau^+ r) }{\langle  r | p_{\sigma_i} | r]} .
}{F_and_G_4d}
Here $\sigma^-$ and $\sigma^+$ are the subsets of $\sigma$ with minus and plus helicity legs kept in the same canonical order as in $\sigma$, and likewise for $\tau^-$ and $\tau^+$ as subsets of $\tau$.  Again, we emphasize that the null reference momenta $q$ and $r$ are arbitrary.

\subsubsection{Nonlinear Sigma Model}

We can also derive a formula for all BCJ numerators in the NLSM by transmuting \cite{transmutation} the BCJ numerators of YM theory.  To accomplish this task we exploit that the amplitudes of YM theory and NLSM are related by the simple kinematic replacement, 
\eq{
p_i \varepsilon_j &\arrowNLSM  \left\{
\begin{array}{ll}
 p_i p_j   &,\quad j\neq n \\ \\
0 & ,\quad  j=n
\end{array} \right. \\  \\
\varepsilon_i \varepsilon_j &\arrowNLSM \left\{
\begin{array}{ll}
0   &,\quad i,j\neq n \\ \\
-\dfrac{p_i q}{p_n q} & ,\quad  j=n
\end{array} \right. ,
}{transmute_Mand}
where $q$ is an arbitrary reference momentum.  The above substitution is a close cousin of the transmutation relations of \cite{transmutation} and is equivalent to the $\otimes$ NLSM replacement rule described earlier, albeit at the level of kinematic invariants.    We opt for this version of transmutation because it treats all legs on equal footing except the root leg.  Applying the transmutation operation in \Eq{transmute_Mand} to \Eq{FandG}, we map the field strength trace and kinematic function to
\eq{
F[\sigma n]  &\arrowNLSM \left\{
\begin{array}{ll}
 {\rm tr}[ \Pi_{\sigma} \widetilde \Pi_n ] &,\quad |\sigma n| \textrm{ even} \\ \\
0 & ,\quad |\sigma n| \textrm{ odd}
\end{array} \right. 
\\ \\
G[\sigma, \tau,\rho n] &\arrowNLSM \left\{
\begin{array}{ll}
- \dfrac{ p_\sigma \Pi_\tau q_{\sigma\rho} }{p_{\sigma\rho n} q_{\sigma \rho}}  &,\quad |\tau| \textrm{ even} \\ \\
0 & ,\quad |\tau| \textrm{ odd}
\end{array} \right.  .
}{F_and_G_NLSM}
Here we have implicitly assumed that the reference momenta in \Eq{FandG} are arbitrary combinations of the external momenta, so they should be treated as such when applying \Eq{transmute_Mand}.
In analogy with \Eqs{Fdef1,Fdef2} we have also defined symmetric momentum tensors,
\eq{
\left[\Pi_i\right]_{\mu\nu} &= p_{i \mu}p_{i \nu}  \qquad \textrm{for} \qquad i\neq n   \\
{}[ \widetilde \Pi_n  ]_{\mu\nu}  &= \frac{q_\mu q_\nu}{(p_n q)^2}  ,
}{Pidef}
as well as the ordered product,
\eq{
\Pi_\sigma = \prod_i^{|\sigma|} \Pi_{\sigma_i}.
}{}
In conclusion, we find that the all NLSM amplitudes are given by \Eqs{result1,result2} with the field strength trace and kinematic function defined in \Eq{F_and_G_NLSM}.
As per the standard double copy prescription, inserting two copies of the BCJ numerators of the NLSM into \Eq{result3} produces the amplitudes of the SG, while inserting one BCJ numerator from YM theory and one from the NLSM generates the amplitudes of BI theory.

Curiously, \Eq{F_and_G_NLSM} is identical to \Eq{FandG} but with field strengths replaced with symmetric momentum tensors.  Since the latter are quadratic in momenta, the resulting BCJ numerators manifestly exhibit an Adler zero condition for all but the root leg.  This creates an elegant parallel between the NLSM and YM theory. In particular, recall that YM theory and GR are uniquely fixed by gauge symmetry, while the NLSM, SG, and BI are uniquely fixed by soft theorems \cite{EFTFromSoft, EFTRecursion, EFTPeriodicTable,JaroslavPion, VectorEFTBootstrap}.  We have derived BCJ numerators for the NLSM and YM theory which manifest these defining properties term by term.


\section{Conclusions}

We have derived a formulation of color-kinematics duality and the double copy implemented at the level of fields and equations of motion.  Our principal insight has been to recast the dynamics in terms of currents and field strengths rather than the usual underlying fields, thus eliminating the intrinsic redundancy incurred by field redefinitions.   
This approach makes color-kinematics duality manifest in the NLSM, and a new structure---covariant color-kinematics duality---manifest in YM theory.  The corresponding kinematic algebras are the diffeomorphism and Lorentz algebras, respectively. In both cases the kinematic current, whose conservation law enforces the appropriate kinematic Jacobi identities, is equal to derivatives of the energy-momentum tensor.

A surprising outcome of our analysis is that YM theory is the covariant double copy of GBAS theory and the $F^3$ theory of field strengths.  Similarly, GR is the covariant double copy of EYM theory and $F^3$ theory.  The physical interpretation of this result is self-evident from the equations of motion and easy to state:  the field strength evolves like a charged biadjoint scalar and spacetime curvature evolves like a gravitating field strength.  This  understanding implies a new decomposition of all tree-level scattering amplitudes in YM theory into those of GBAS theory.  From this representation we derive a closed-form, analytic expression for all tree-level BCJ numerators in the NLSM and YM theory for any number of external legs in arbitrary spacetime dimension.  Via the standard double copy procedure, these constitute explicit formulas for all tree-level amplitudes in YM, GR, NLSM, SG, and BI.

The present analysis leaves numerous avenues for future inquiry.  First and foremost is the question of loops.  Obviously, we have articulated all of our results squarely in terms of classical equations of motion and tree-level scattering.  However, it may be feasible to generalize our approach to include radiative corrections.  If so, a key step will be to reframe the dynamics in terms of currents and field strengths where possible.

Also worthy of study is the topic of higher-dimension operators.  Previous works \cite{LanceF3, ElvangPhi3, PencoHigherDim} have explored the  corrections of YM theory  and the NLSM which are compatible with color-kinematics duality.   Following \cite{JJHigherDim}, it would be interesting to implement the double copy---at the level of fields  and equations of motion---to those higher-dimension operators in BAS and GBAS theory to derive the corresponding corrections in the NLSM and YM theory. 

The present analysis has focused entirely on those theories directly linked to YM theory and the NLSM via the double copy.  Nevertheless, it may certainly be possible to extend these results to a broader web of theories, {\it e.g.}~including supersymmetry.  Another natural question is whether there exist new covariant double copy constructions based on the product of the $F^3$ theory with other gauged variations of BAS theory and the NLSM.

Last but not least is the question of color-kinematics duality beyond flat space, which has enjoyed a recent resurgence of interest \cite{AdS1, AdS2, AdS3, AdS4, AdS5}.   Since our approach leans heavily on equations of motion, it seems straightforward to lift our framework to curved spacetime.  However, once the natural observables are off-shell correlators, it remains to be seen how much can be accomplished without the aid of on-shell conditions which were so crucial for our analysis.

 \begin{center} {\bf Acknowledgments}
\end{center}
\noindent 
We are grateful to Zvi Bern, J.J.~Carrasco, Lance Dixon, Andreas Helset, Julio Parra-Martinez, Radu Roiban, Ira Rothstein, and Mikhail Solon
for useful discussions and comments on the paper.
C.C.~and J.M.~are supported by the DOE under grant no.~DE-SC0011632 and by the Walter Burke Institute for Theoretical Physics. 

\appendix

\section{Alternative Formulations}

\subsection{Nonlinear Sigma Model}

\label{app:twoform}

There is another representation of the NLSM that also manifests color-kinematics duality.
This formulation has the advantage that the leaf and root legs are treated more equitably.  
To begin, let us introduce an adjoint antisymmetric tensor field $j^a_{\mu\nu}$ that satisfies the Bianchi identity,
\eq{
\partial_{[\rho} j_{\mu\nu]}^a=0.
}{Bianchi_twoform}
Furthermore, we demand that $j^a_{\mu\nu}$ is related to the chiral current $j^a_\mu$ by
\eq{
\partial^\mu j_{\mu\nu}^a = j_\nu^a - \frac{q_\nu}{q\partial} J^a.
}{twoform_relation}
By construction, the divergence of the above equation, $\partial^\nu \left[\textrm{\Eq{twoform_relation}}\right]_{\nu }  $, exactly reproduces the NLSM equation of motion in \Eq{EOM_NLSM2}.

Using \Eq{Bianchi_twoform} and \Eq{twoform_relation}, we substitute $j^a_{\mu\nu}$ for $j^a_\mu$ into the pure gauge condition of \Eq{EOM_NLSM1} to obtain a new formulation of the NLSM in terms of an equation of motion for the antisymmetric tensor field,
\eq{
\Box j^a_{\mu\nu} + f^{abc} \partial^\rho j^b_{\rho\mu} \partial^\sigma j^c_{\sigma\nu} = \frac{\partial_{[\mu} q_{\nu]}}{q\partial} J^a.
}{EOM_NLSM_twoform}
We do not present the Feynman rules for this theory here but they are trivially obtained from the above equations of motion in the usual way.   In this representation the polarizations of the leaf legs and the root leg are the same, placing them on the same footing.

As before, we can again read off the structure of the kinematic algebra by directly comparing \Eq{EOM_NLSM_twoform} against \Eq{EOM_BAS}.  In this alternative formulation, the $\otimes$ NLSM replacement rules of \Eq{replace_NLSM} become
 \eq{
{\cal V}^{ a} \quad &\arrowNLSM \quad {\cal V}_{\mu\nu} \\
f^{ a  b  c} {\cal V}^{b} {\cal W}^{ c} \quad &\arrowNLSM \quad 
\partial^\rho {\cal V}_{\rho\mu} \partial^\sigma {\cal W}_{\sigma\nu} - \partial^\rho {\cal W}_{\rho\mu} \partial^\sigma {\cal V}_{\sigma\nu} \\
J^{a} \quad &\arrowNLSM \quad \frac{\partial_{[\mu} q_{\nu]}}{q\partial} J.
}{replace_NLSM_twoform}
Applying these substitutions to the dual color current of BAS theory, we obtain
\eq{
{\cal K}_{\mu\nu\alpha}^{\NLSM} =  \partial^\rho j^a_{\rho\mu}  \overset{\leftrightarrow}{\partial}_\alpha \partial^\sigma j^a_{\sigma\nu},
}{K_NLSM_twoform}
which is the kinematic current for this alternative formulation of the NLSM.  This current is conserved on the support of the equation of motion in \Eq{EOM_NLSM_twoform}.

\subsection{Yang-Mills Theory}

\label{app:cubic}

It is possible reformulate YM theory purely in terms of a dynamical field strength that self-interacts via a single cubic vertex.  Taking the particular combination of {\it partial} derivatives, $\partial^\rho \left[\textrm{\Eq{EOM_YM1}}\right]_{\rho\mu\nu} + \partial_{[\mu} \left[\textrm{\Eq{EOM_YM2}}\right]_{\nu]}  $, we obtain \eq{
\Box F^{a}_{\mu\nu} +f^{abc} \left[ \partial^\rho (A^b_{[\rho} F^c_{\mu\nu]} ) - \partial_{[\mu} (A^{b \rho} F^{c}_{\nu  ] \rho}) \right] &= \partial_{[\mu} J_{\nu]}^{a}  \qquad \textrm{where} \qquad A^a_\mu = - \frac{q^\nu F_{\mu \nu}^a}{q\partial} .
}{EOM_YM_alt}
By choosing the axial gauge condition from \Eq{A_from_F}, we have mandated that the gauge field is linearly related to the field strength.  Hence we can eliminate the gauge field altogether, and \Eq{EOM_YM_alt} is actually an equation of motion for the field strength {\it alone}.  Furthermore, this field strength exhibits a single cubic self-interaction.

It is straightforward to derive Feynman rules from \Eq{EOM_YM_alt} in the usual way.  The propagator is the same as in \Eq{prop_F3}, while the polarizations for the leaf and root legs are the same as in \Eqs{leaf_pol_F3,root_pol_F3}.  Meanwhile, the cubic Feynman vertex is
\eq{
\raisebox{0ex}{\includegraphics[trim={0 0 0 0},clip,valign=c,scale=0.8]{figs/YMVert}} & = 
\left[\frac{i f^{a_1 a_2 a_3}q^{\mu_1}\eta^{\mu_2\mu_3}}{4qp_1}  \left( \frac{1}{2} p_3^{\nu_1} \eta^{\nu_2\nu_3} - p_3^{\nu_2}\eta^{\nu_3\nu_1} + p_3^{\nu_3} \eta^{\nu_1\nu_2} \right) \right]_{\rm antisym}  + \{ 1 \leftrightarrow 2 \} ,
}{Vert_F3app}
where the expression in square brackets is separately antisymmetrized over each pair of spacetime indices, $\mu_1 \nu_1$, $\mu_2 \nu_2$, and $\mu_3 \nu_3$, sans additional numerical normalization factors.  The symmetrization on legs 1 and 2 simply swaps all color and kinematic labels for those states.    Much like our other formulations of the NLSM, this representation of YM theory only manifests permutation invariance on the leaf legs. 

We have verified that the above Feynman rules correctly reproduce the amplitudes of YM theory through six-point scattering.
However, the contributing Feynman diagrams do not manifest color-kinematics duality, {\it i.e.}~they do not automatically satisfy the kinematic Jacobi relations.   As is well-known \cite{BCJReview}, one can still obtain a valid double copy from the BCJ product of two amplitudes provided {\it at least one of them} satisfies the kinematic Jacobi identities.  Thus, by taking the product of the Feynman rules for YM described above, together with those of the NLSM, one obtains a new cubic formulation of BI theory.

It is striking that the entirety of YM theory is encoded in a single cubic interaction of the field strength.  It would be interesting to see whether this formulation offers any computational advantages over existing implementations of Berends-Giele recursion for YM theory \cite{BerendsGiele}.

\section{Fundamental BCJ Relation}

\label{app:BCJ}

The fundamental BCJ relation \cite{BCJ1} can actually be derived directly from equations of motion.   The key ingredient in this approach follows from \cite{Brown:2016mrh}, whose authors discovered an intriguing ``color factor symmetry'' of on-shell scattering which enforces the fundamental BCJ relation.
Our only insight here is to realize that this color factor symmetry is actually an invariance of the equations of motion themselves. 

 For concreteness,  consider the case of BAS theory.   The color factor symmetry is defined by an infinitesimal shift of the color structure constant,
\eq{
f^{ a b c} \rightarrow f^{ a b c} + \delta f^{ a b c},
}{CFS}
by a perturbation that induces {\it kinematic} dependence when contracted with fields, so
\eq{
\delta f^{a b c} {\cal V}^{b} {\cal W}^{ c}  = \epsilon^{a} \delta^{  b c}  {\cal V}^b \overset{\leftrightarrow}{\Box} {\cal W}^c,
}{}
for an arbitrary reference color vector  $\epsilon^a$.   This perturbation preserves permutation invariance on the leaf legs but not the root leg.
Applying \Eq{CFS} to the BAS equations of motion in \Eq{EOM_BAS}, we see that the kinetic term is trivially invariant while the interaction term shifts by
\eq{
\delta f^{a b c} f^{\ov a \ov b \ov c} \phi^{b \ov b} \phi^{c\ov c} 
= \epsilon^a \partial^\alpha {\cal K}_\alpha^{\ov a} =0,
}{}
which vanishes on-shell on account of conservation of the dual color current in \Eq{JK_BAS}.   

Hence, the equations of motion are invariant under the color factor symmetry.  As an immediate corollary, any perturbative solution to the equations of motion---and thus any on-shell scattering amplitude---is also invariant.  Indeed,  in terms of the cubic Feynman vertex of BAS theory in \Eq{vert_BAS}, the perturbation of the color structure constant becomes
\eq{
\delta f^{a_1 a_2 a_3} = \epsilon^{a_3} \delta^{a_1 a_2}  (p_1^2 - p_2^2),
}{color_shift}
which coincides exactly with the color factor symmetry of \cite{Brown:2016mrh} when the root leg is on-shell.

To derive the fundamental BCJ relation we simply repeat the strategy of \cite{Brown:2016mrh}.
In particular, consider an amplitude of BAS theory expressed in the half ladder representation of \cite{DDM}.  Color factor symmetry implies that this object should be invariant under the transformation in \Eq{CFS}.
With the benefit of hindsight, we choose the color reference vector in \Eq{color_shift} to be orthogonal to {\it all color indices} in the amplitude except for that of a single external leg of our choice.  For our purposes this special leg can be any external leg that is not at either end of the half ladders.  In the resulting shift of the amplitude, each term includes a color half ladder with the special leg removed, multiplying the difference of the inverse propagators on either side of that leg, as dictated by \Eq{color_shift}.  Setting the coefficient of each independent color structure to zero, we obtain the fundamental BCJ relation.


The above discussion has centered purely on BAS theory, but our manipulations apply to any theory whose equations of motion manifest color-kinematics duality and whose Feynman rules automatically satisfy the kinematic Jacobi identities.

\bibliographystyle{utphys-modified}
\bibliography{CCKbib}

\end{document}